\documentclass[12pt]{JHEP3}
\usepackage{axodraw}

\def\Year{\expandafter\eatPrefix\the\year}
\newcount\hours \newcount\minutes
\def\monthname{\ifcase\month\or
January\or February\or March\or April\or May\or June\or July\or
August\or September\or October\or November\or December\fi}
\def\shortmonthname{\ifcase\month\orx
Jan\or Feb\or Mar\or Apr\or May\or Jun\or Jul\or
Aug\or Sep\or Oct\or Nov\or Dec\fi}

\def\TimeStamp{\hours\the\time\divide\hours by60%
\minutes -\the\time\divide\minutes by60\multiply\minutes by60%
\advance\minutes by\the\time
${\rm \shortmonthname}\cdot   \if\day<10{}0\fi\the\day\cdot   \the\year
\qquad\the\hours:\if\minutes<10{}0\fi\the\minutes$}

\newskip\humongous \humongous=0pt plus 1000pt minus 100pt
\def\caja{\mathsurround=0pt}
\def\eqalign#1{\,\vcenter{\openup1\jot \caja
       \ialign{\strut \hfil$\displaystyle{##}$&$
        \displaystyle{{}##}$\hfil\crcr#1\crcr}}\,}
\newif\ifdtup

\newcounter{eqnumber}[section]
\renewcommand{\theeqnumber}{\thesection.\arabic{eqnumber}}
\def\equn{\refstepcounter{eqnumber}
\eqno({\rm \theeqnumber})
}

\def\npb#1#2#3{{\rm Nucl. Phys. B}{\bf \ #1}, #3 (#2)}
\def\plb#1#2#3{{\rm Phys. Lett. B}{\bf \ #1}, #3 (#2)}

\def\cqg#1#2#3{{\rm Class. and Quant.\ Grav.} {\bf  #1}, #3 (#2)}

\def\hepth#1{[hep-th/#1]}
\def\hepph#1{[hep-ph/#1]}

\def\Fn{}
\def\Fs#1#2{F^{{#1}}_{\Fn #2}}
\def\Fone{\Fs{\rm 1m}}
\def\Feasy{\Fs{{\rm 2m}\,e}}
\def\Fhard{\Fs{{\rm 2m}\,h}}
\def\Fthree{\Fs{\rm 3m}}
\def\Ffour{\Fs{\rm 4m}}
\def\Wsix#1{W_6^{(#1)}}
\def\cc{\dagger}

\newbox\charbox
\newbox\slabox
\def\s#1{{      
        \setbox\charbox=\hbox{$#1$}
        \setbox\slabox=\hbox{$/$}
        \dimen\charbox=\ht\slabox
        \advance\dimen\charbox by -\dp\slabox
        \advance\dimen\charbox by -\ht\charbox
        \advance\dimen\charbox by \dp\charbox
        \divide\dimen\charbox by 2
        \raise-\dimen\charbox\hbox to \wd\charbox{\hss/\hss}
        \llap{$#1$}
}}

\def\spa#1.#2{\left\langle#1\,#2\right\rangle}
\def\spb#1.#2{\left[#1\,#2\right]}
\def\lor#1.#2{\left(#1\,#2\right)}

\catcode`@=11  

\def\Slash#1{\hskip 0.05 cm \slash\hskip -0.22 cm #1}

\def\Tr{\, {\rm Tr}}

\def\eps{\epsilon}

\def\e{\epsilon}

\def\half{{1\over 2}}

\def\la{\langle}
\def\ra{\rangle}

\def\lsl{\not{\hbox{\kern-2.3pt $\ell$}}}
\def\ksl{\not{\hbox{\kern-2.3pt $k$}}}

\def\Z{0}

\def\rg{c_{\Gamma}}

\def\spa#1.#2{\left\langle#1\,#2\right\rangle}
\def\spb#1.#2{\left[#1\,#2\right]}
\def\lor#1.#2{\left(#1\,#2\right)}

\def\sand#1.#2.#3{%
  \left\langle\smash{#1}{\vphantom1}\right|{#2}%
  \left|\smash{#3}{\vphantom1}\right\rangle}
\def\sandp#1.#2.#3{%
  \left\langle\smash{#1}{\vphantom1}^{-}\right|{#2}%
  \left|\smash{#3}{\vphantom1}^{+}\right\rangle}
\def\sandpp#1.#2.#3{%
  \left\langle\smash{#1}{\vphantom1}^{+}\right|{#2}%
  \left|\smash{#3}{\vphantom1}^{+}\right\rangle}
\def\sandmm#1.#2.#3{%
  \left\langle\smash{#1}{\vphantom1}^{-}\right|{#2}%
  \left|\smash{#3}{\vphantom1}^{-}\right\rangle}
\def\sandpm#1.#2.#3{%
  \left\langle\smash{#1}{\vphantom1}^{+}\right|{#2}%
  \left|\smash{#3}{\vphantom1}^{-}\right\rangle}
\def\sandmp#1.#2.#3{%
  \left\langle\smash{#1}{\vphantom1}^{-}\right|{#2}%
  \left|\smash{#3}{\vphantom1}^{+}\right\rangle}

\def\Atree{A^{\rm tree}}

\def\A#1{{\cal A}_{#1}}

\def\L{\left(}\def\R{\right)}

\def\LB{\left[}\def\RB{\right]}

\def\tn#1#2{t^{[#1]}_{#2}}
\def\L{\left(}\def\R{\right)}

\def\BR#1#2{\la#1^+|\Slash{K}|#2^+\ra}

\def\BRi#1#2#3{\la#1|{K}_{#3}|#2\ra}
\def\tree{{\rm tree}}

\def\Gr{{\rm Gr}}

\def\NeqEight{{\cal N} = 8}
\def\NeqFour{{\cal N} = 4}
\def\NeqOne{{\cal N} = 1}
\def\NeqZero{{\cal N} = 0}
\def\NeqTwo{{\cal N} = 2}

\def\gluino{\Lambda}
\def\gluinb{\bar\Lambda}

\def\Ang{A_{n}}

\def\Ant{A_{n}^{\rm tr}}
\def\Ast{A_{6}^{\rm tr}}

\def\Asf{A_{6}^{\NeqFour}}
\def\Anz{A_{n}^{\NeqTwo}}
\def\Asz{A_{6}^{\NeqTwo}}

\title{Supersymmetric Ward Identities and NMHV Amplitudes involving
Gluinos}
\author{Steven J. Bidder, 
David~C.~Dunbar
and Warren B. Perkins \\
Department of Physics \\
University of Wales Swansea, Swansea, SA2 8PP, UK}

\preprint{SWAT-05-430}

\abstract{ 
We show how Supersymmetric Ward Identities can be used to
obtain amplitudes involving gluinos or adjoint scalars from
purely gluonic amplitudes. We obtain results for all
one-loop six-point NMHV amplitudes in $\NeqFour$ Super Yang-Mills
theory which involve two gluinos or two scalar particles.
More general cases are also discussed. 
}


\begin{document}

\section{Introduction}

Recently, inspired by a possible duality between gauge theory and twistor
string theory~\cite{Witten:2003nn,CSW}, there has been much progress in
obtaining one-loop gauge theory amplitudes in compact
forms~\cite{BDDK7,BDKn,BrittoUnitarity,Cachazo:2004dr,Britto:2004nj,Bidder:2004tx,BBDP2005a,BrittoSQCD}.  
Most of
the applications in loop calculations have been to amplitudes which
involve only external gluons.  In this paper we explore how amplitudes
involving particles other than gluons may be obtained via symmetry
constraints rather than by direct computation.  In particular, we use
Supersymmetric Ward Identities~\cite{SWI}(SWI) to obtain amplitudes
involving gluinos and scalars from purely gluonic amplitudes.

On-shell SWI~\cite{SWI} impose
powerful constraints on amplitudes in gauge theories,  giving
algebraic constraints between amplitudes with the same helicity
configuration but different external particle types.  These
constraints apply at any order in perturbation theory.
From a Feynman diagram perspective, these relationships are most
naturally employed to obtain purely gluonic amplitudes from amplitudes
involving fermions, as the latter are easier to calculate. For
example, at six-point, Kunszt~\cite{Zoltan} was able to obtain the
purely gluonic tree amplitudes from the set of amplitudes with four
gluons and two fermions.  Motivated by the recent advances in
calculating purely gluonic amplitudes, in this paper we will reverse
this process and generate amplitudes involving fermions from the
purely gluonic ones.  For some helicity configurations the SWI contain
sufficient information to simply solve for the fermionic amplitudes:
for example in $\NeqFour$ gauge theory the SWI
 for amplitudes with two negative helicities and the rest
positive (known as MHV amplitudes) can be easily solved and amplitudes
with any external particles obtained from the purely gluonic MHV
amplitudes~\cite{ParkeTaylor,Nair} by a simple multiplicative factor.

For other configurations, such as those with three negative helicities
(known as ``next-to-MHV'' or NMHV amplitudes) the SWI do not allow
such simple solutions. However, we shall show how the SWI can be
solved in a natural way to obtain amplitudes with two gluinos in terms
of the purely gluonic case.  We will first apply this to the six-point
tree amplitudes where we can connect to known expressions.  Secondly
we shall determine the one-loop six-point NMHV amplitudes in
$\NeqFour$ SYM which involve two gluinos.  More generally there also exist SWI which
involve amplitudes with two gluinos, four gluinos, two scalars and two gluinos plus a scalar.  
We explicitly determine the two scalar amplitudes. The SWI then give the remaining amplitudes
directly in terms of known amplitudes.

\section{$\NeqFour$  Amplitudes and Recent Developments}

In this section, we describe the organisation of tree and one-loop
amplitudes and review the recent progress in determining the one-loop
gluonic amplitudes in $\NeqFour$ SYM.

\vskip 15 pt 
\noindent
{\bf Tree Amplitudes: }
Tree-level amplitudes for $U(N_c)$ or $SU(N_c)$ gauge theories with
$n$ external adjoint particles can be decomposed into colour-ordered partial
amplitudes multiplied by an associated
colour-trace~\cite{TreeColour,ManganoReview}.  Summing over
all non-cyclic permutations reconstructs the full amplitude
$\A{n}^\tree$ from the partial amplitudes $A_n^\tree(\sigma)$,
$$
\A{n}^\tree(\{k_i,a_i\}) =
g^{n-2} \sum_{\sigma\in S_n/Z_n} \Tr(T^{a_{\sigma(1)}}
\cdots T^{a_{\sigma(n)}})
\ A_n^\tree(k_{\sigma(1)}
,\ldots,
            k_{\sigma(n)})
\ ,
\equn\label{TreeAmplitudeDecomposition}
$$
where $k_i$ and $a_i$ are respectively the momentum
and colour-index of the $i$-th external
particle, $g$ is the coupling constant and $S_n/Z_n$ is the set of
non-cyclic permutations of $\{1,\ldots, n\}$.
The $U(N_c)$ ($SU(N_c)$) generators $T^a$ are the set of
traceless hermitian $N_c\times N_c$ matrices,
normalised such that $\Tr\L T^a T^b\R = \delta^{ab}$.
Conventionally we take all particles to be outgoing. We denote gluons 
by $g_i$ and 
adjoint fermions by 
$\Lambda_i$. We will often refer to the adjoint fermions as gluinos for simplicity. 

Amplitudes involving fundamental particles, for example fermions (or quarks) 
$\lambda_i$ have a different decomposition~\cite{TreeColour},
$$
\A{n}^\tree(\bar\lambda_1,\lambda_2,g_3,\cdots, ) =
g^{n-2} \sum_{\sigma\in S_{n-2}} 
\left(  T^{a_{\sigma(3)}}
\cdots T^{a_{\sigma(n)}}\right)_{i_2}{}^{\bar i_1} 
\ A_n^\tree( \bar\lambda_1,\lambda_2, g_{\sigma(3)}
,\ldots,
            g_{\sigma(n)}
)\ ,
\equn\label{TreeQuarkDecomposition}
$$ where $i_2$ and $\bar i_1$ are the colour indices on the quarks. 
Note that the two quarks are adjacent in the ordering. 
The partial tree amplitudes for two quarks are simply related to those of 
adjoint fermions by,
$$ 
A_n^\tree( \bar\lambda_1^+,\lambda_2^-, g_3^{ }, g_4^{ } \cdots
g_n^{ } ) 
=A_n^\tree( \bar\gluino_1^+,\gluino_2^-, g_3^{ }, g_4^{ } \cdots
g_n^{ } ) ,
\equn
$$ 
and
the difference between the full amplitudes is entirely in
the colour factors.
For amplitudes where exactly two of the external particles are quarks or gluinos and the remainder
are gluons, the amplitude will vanish unless the quarks or gluinos have opposite
helicity.

Tree amplitudes where all the particles have the same helicity
vanish, as do amplitudes where all but one of the helicities  are identical,
$$
A_n^\tree(1^{\pm},2^{+}, \ldots,n^+) = 0.
\equn
$$
The simplest
non-vanishing amplitudes are the MHV amplitudes with two particles of negative 
helicity and the remainder positive. 
The MHV partial amplitudes for gluons are given by the 
Parke-Taylor formulae~\cite{ParkeTaylor}, 
$$
\eqalign{
  \Atree_n(g_1^+,\ldots,g_j^-,\ldots,g_k^-,
                \ldots,g_n^+),
&=\ i\, { {\spa{j}.{k}}^4 \over \spa1.2\spa2.3\cdots\spa{n}.1 }\ ,
  \cr}
\equn\label{ParkeTaylor}
$$ for a partial amplitude where $j$ and $k$ are the legs with
negative helicity.  We use the notation  
$\spa{j}.{l}\equiv \langle j^- | l^+ \rangle $, 
$\spb{j}.{l} \equiv \langle j^+ |l^- \rangle $, 
with $| i^{\pm}\ra $ 
being a massless Weyl spinor with
momentum $k_i$ and chirality
$\pm$~\cite{SpinorHelicity,ManganoReview}.  The spinor products are
related to momentum invariants by 
$\spa{i}.j\spb{j}.i=2k_i \cdot k_j\equiv s_{ij}$ 
with $\spa{i}.j^*=\spb{j}.i$.

The MHV amplitudes with external particles other
than gluons can be obtained from these using SWI. 
Formulae linking these amplitudes can also be derived using current algebra
techniques~\cite{Nair}. Formulae for amplitudes with three minus
helicity gluons can be deduced by recursion relations but are
more complicated~\cite{Kosower:1989xy}.

\vskip 15 pt 
\noindent
{\bf One-Loop Amplitudes: }
For one-loop amplitudes of adjoint particles, 
one may perform a colour decomposition similar
to the tree-level decomposition~(\ref{TreeAmplitudeDecomposition})~\cite{Colour}.
In this case there are  two traces
over colour matrices
and
the result takes the form,
$$
{\cal A}_n^{\rm 1-loop}\L \{k_i,a_i\}\R =
g^n \,\sum_{c=1}^{\lfloor{n/2}\rfloor+1}
      \sum_{\sigma \in S_n/S_{n;c}}
     \Gr_{n;c}\L \sigma \R\,A_{n;c}^{}(\sigma),
\label{ColourDecomposition}\equn
$$
where ${\lfloor{x}\rfloor}$ is the largest integer less than or equal to $x$.
The leading colour-structure factor,
$$
\Gr_{n;1}(1) = N_c\ \Tr\L T^{a_1}\cdots T^{a_n}\R \, ,\equn
$$
is just $N_c$ times the tree colour factor, and the subleading colour
structures ($c>1)$ are given by,
$$
\Gr_{n;c}(1) = \Tr\L T^{a_1}\cdots T^{a_{c-1}}\R\,
\Tr\L T^{a_c}\cdots T^{a_n}\R \, .\equn
$$
$S_n$ is the set of all permutations of $n$ objects
and $S_{n;c}$ is the subset leaving $\Gr_{n;c}$ invariant.
Once again it is convenient to use $U(N_c)$ matrices; the extra $U(1)$
decouples~\cite{Colour}.
                                                                               
For one-loop amplitudes the subleading in colour amplitudes
$A_{n;c}$,   $c > 1 $,
may be obtained from summations of permutations
of the leading in colour amplitude~\cite{BDDKa},
$$
 A_{n;c}(1,2,\ldots,c-1;c,c+1,\ldots,n)\ =\
 (-1)^{c-1} \sum_{\sigma\in COP\{\alpha\}\{\beta\}} A_{n;1}(\sigma),
\equn
$$
where $\alpha_i \in \{\alpha\} \equiv \{c-1,c-2,\ldots,2,1\}$,
$\beta_i \in \{\beta\} \equiv \{c,c+1,\ldots,n-1,n\}$,
and $COP\{\alpha\}\{\beta\}$ is the set of all
permutations of $\{1,2,\ldots,n\}$ with $n$ held fixed
that preserve the cyclic
ordering of the $\alpha_i$ within $\{\alpha\}$ and of the $\beta_i$
within $\{\beta\}$, while allowing for all possible relative orderings
of the $\alpha_i$ with respect to the $\beta_i$.
Hence, we need only focus on the leading in colour amplitude $A_{n;1}$ 
(which we will generally abbreviate to $A_n$) 
and use this
relationship to generate the full amplitude if required.

One-loop amplitudes depend on the particles circulating within the
loop and thus on the spectrum of the theory.  In supersymmetric
amplitudes there are generically cancellations between the bosons and
fermions in the loop.  For $\NeqFour$ SYM these cancellations lead to 
considerable simplifications in the loop momentum integrals. This is
manifest in the ``string-based approach'' to computing loop
amplitudes~\cite{StringBased}.  As a result of these simplifications,
$\NeqFour$ one-loop amplitudes can be expressed simply as a sum of
scalar box-integral functions~\cite{BDDKa},
 $$
  I_{i}^{1{\rm m}}
\hskip 0.5truecm
 I_{r;i}^{2{\rm m}e}
\hskip 0.5truecm
  I_{r;i}^{2{\rm m}h}
\hskip 0.5truecm
  I_{r,r',i}^{3{\rm m}}
\hskip 0.5truecm
I_{r, r', r'', i}^{4{\rm m}} \equn
$$
with the labeling as indicated,
\begin{center}
\begin{picture}(120,95)(0,0)
\Line(30,20)(70,20)
\Line(30,60)(70,60)
\Line(30,20)(30,60)
\Line(70,60)(70,20)
                                                                                
\Line(30,20)(15,5)
\Line(30,60)(15,75)
\Line(70,20)(85,5)
\Line(70,60)(85,75)
                                                                                
\Line(30,20)(30,5)
\Line(30,20)(15,20)
\Text(22,5)[c]{\small$\bullet $}
\Text(15,12)[c]{\small $\bullet $}
                                                                                
\Text(32,5)[l]{\small $\hbox{\rm i}$}
\Text(85,10)[l]{\small $\hbox{\rm i-1}$}
\Text(85,70)[l]{\small $\hbox{\rm i-2}$}
\Text(25,70)[l]{\small $\hbox{\rm i-3}$}

\Text(40,40)[l]{$I^{1m}_{i}$}
\end{picture}
\begin{picture}(120,95)(0,0)
\Line(30,20)(70,20)
\Line(30,60)(70,60)
\Line(30,20)(30,60)
\Line(70,60)(70,20)
                                                                                
\Line(30,20)(15,5)
\Line(30,60)(15,75)
\Line(70,20)(85,5)
\Line(70,60)(85,75)
                                                                                
\Line(30,20)(30,5)
\Line(30,20)(15,20)
\Text(22,5)[c]{\small$\bullet$}
\Text(15,12)[c]{\small $\bullet$}
                                                                                
\Line(70,60)(70,75)
\Line(70,60)(85,60)
\Text(75,75)[c]{\small$\bullet$}
\Text(85,65)[c]{\small $\bullet$}

\Text(32,5)[l]{\small $\hbox{\rm i}$}
\Text(85,10)[l]{\small $\hbox{\rm i-1}$}
\Text(25,70)[l]{\small $\hbox{\rm i+r}$}

\Text(40,40)[l]{$I^{2me}_{r;i}$}
\end{picture}
\begin{picture}(120,95)(0,0)
\Line(30,20)(70,20)
\Line(30,60)(70,60)
\Line(30,20)(30,60)
\Line(70,60)(70,20)
                                                                                
\Line(30,20)(15,5)
\Line(30,60)(15,75)
\Line(70,20)(85,5)
\Line(70,60)(85,75)
                                                                                
\Line(30,20)(30,5)
\Line(30,20)(15,20)
\Text(22,5)[c]{\small$\bullet$}
\Text(15,12)[c]{\small $\bullet$}
                                                                                
\Line(30,60)(30,75)
\Line(30,60)(15,60)
\Text(15,65)[c]{\small$\bullet$}
\Text(25,75)[c]{\small $\bullet$}

\Text(32,5)[l]{\small $\hbox{\rm i}$}
\Text(85,10)[l]{\small $\hbox{\rm i-1}$}
\Text(85,70)[l]{\small $\hbox{\rm i-2}$}
\Text(5,55)[l]{\small $\hbox{\rm i+r}$}

\Text(40,40)[l]{$I^{2mh}_{r;i}$}
\end{picture}

\begin{picture}(120,95)(0,0)
\Line(30,20)(70,20)
\Line(30,60)(70,60)
\Line(30,20)(30,60)
\Line(70,60)(70,20)
                                                                                
\Line(30,20)(15,5)
\Line(30,60)(15,75)
\Line(70,20)(85,5)
\Line(70,60)(85,75)
                                                                                
\Line(30,20)(30,5)
\Line(30,20)(15,20)
\Text(22,5)[c]{\small $\bullet$}
\Text(15,12)[c]{\small $\bullet$}
                                                                                
\Line(30,60)(30,75)
\Line(30,60)(15,60)
\Text(15,65)[c]{\small $\bullet$}
\Text(25,75)[c]{\small $\bullet$}

\Line(70,60)(70,75)
\Line(70,60)(85,60)
\Text(75,75)[c]{\small $\bullet$}
\Text(85,65)[c]{\small $\bullet$}
                                                                                                                                                                
\Text(32,5)[l]{\small $\hbox{\rm i}$}
\Text(85,10)[l]{\small $\hbox{\rm i-1}$}
\Text(45,80)[l]{\small $\hbox{\rm i+r+r}'$}
\Text(5,55)[l]{\small $\hbox{\rm i+r}$}

\Text(37,40)[l]{$I^{3m}_{r,r',i}$}
\end{picture}
\begin{picture}(120,95)(0,0)
\Line(30,20)(70,20)
\Line(30,60)(70,60)
\Line(30,20)(30,60)
\Line(70,60)(70,20)
                                                                                
\Line(30,20)(15,5)
\Line(30,60)(15,75)
\Line(70,20)(85,5)
\Line(70,60)(85,75)
                                                                                
\Line(30,20)(30,5)
\Line(30,20)(15,20)
\Text(22,5)[c]{\small $\bullet$}
\Text(15,12)[c]{\small $\bullet$}
                                                                                
\Line(30,60)(30,75)
\Line(30,60)(15,60)
\Text(15,65)[c]{\small $\bullet$}
\Text(25,75)[c]{\small $\bullet$}
                                                                                                                                                                
\Line(70,60)(70,75)
\Line(70,60)(85,60)
\Text(75,75)[c]{\small $\bullet$}
\Text(85,65)[c]{\small $\bullet$}
                                                                                
\Line(70,20)(70,5)
\Line(70,20)(85,20)
\Text(75,5)[c]{\small $\bullet$}
\Text(85,15)[c]{\small $\bullet$}

\Text(32,5)[l]{\small $\hbox{\rm i}$}
\Text(85,27)[l]{\small $\hbox{\rm i+r+r}'\hbox{\rm +r}''$}
\Text(45,80)[l]{\small $\hbox{\rm i+r+r}'$}
\Text(5,55)[l]{\small $\hbox{\rm i+r}$}

\Text(30,40)[l]{$I^{4m}_{r,r'\hskip -2pt ,r''\hskip -2pt,i}$}
\end{picture}

\end{center}
                                                                                
Explicit forms for these scalar box integrals can be found in ref.~\cite{BDDKa}. 
The four
dimensional boxes have dimension $-2$. It is convenient to
define dimension zero $F$-functions by removing the momentum
prefactors of the $D=4$ scalar boxes~\cite{BDDKb},
$$
\eqalign{
&  I_{i}^{1{\rm m}} = -2 \rg {\Fone{i} \over \tn{2}{i-3} \tn{2}{i-2} }
        \,  ,\hskip 0.2 cm
 I_{r;i}^{2{\rm m}e}
= -2 \rg {\Feasy{r;i}
      \over \tn{r+1}{i-1}\tn{r+1}{i} -\tn{r}{i}\tn{n-r-2}{i+r+1} }\,,
        \,  \hskip 0.2 cm
  I_{r;i}^{2{\rm m}h}
= -2 \rg {\Fhard{r;i} \over \tn{2}{i-2} \tn{r+1}{i-1} } \,,
\cr
& \hskip 1.0 truecm   I_{r,r',i}^{3{\rm m}}
= -2 \rg {\Fthree{r,r';i}
     \over \tn{r+1}{i-1} \tn{r+r'}i -\tn{r}{i} \tn{n-r-r'-1}{i+r+r'} }\,,
        \,  \hskip 0.5 cm
I_{ r, r', r'', i}^{4{\rm m}}  =
-2 {\Ffour{r, r', r'';i}\over t_i^{[r+ r']}\; t_{i+r}^{[r'+r'']}\;\rho}\, ,
\cr}\equn
$$
where, 
$$
t_{a}^{[p]}\equiv (k_a+k_{a+1}+\cdots +k_{a+p-1})^2
\ , 
\equn
$$
and,
$$
\rg\ =\ {\Gamma(1+\epsilon)\Gamma^2(1-\epsilon)\over 
          (4\pi)^{2-\epsilon}\Gamma(1-2\epsilon)}.
\equn
$$
The one-loop amplitudes can thus be  expressed as, 
$$
A^{\NeqFour}  = \sum_{i}  c_i F_i \, ,
\equn\label{generalform2}
$$
and the computation of one-loop $\NeqFour$ amplitudes is then just a matter of 
determining the rational coefficients $c_i$.
These remarkable simplifications also appear to extend beyond one-loop~\cite{MultiLoop}. 

Furthermore, it has been shown that these amplitudes are ``cut-constructible'',
in that the coefficients can be determined from unitary cuts. 
Using this fact, in ref~\cite{BDDKa} the one-loop amplitudes 
were determined for the all-$n$
MHV amplitudes and in ~\cite{BDDKb} the remaining six-point gluonic amplitudes
(the NMHV amplitudes) were computed and the MHV amplitudes determined in 
$\NeqOne$ theories. 

The amplitude for gluonic scattering in $\NeqFour$ theory can be
thought of as a component of gluonic scattering in non-supersymmetric
theories. One can decompose the one-loop pure gluon amplitude as a sum
of contributions from matter supermultiplets,
$$
A_{n}\ \equiv\
A_{n}^{\;\NeqFour}-4A_{n}^{\;\NeqOne\ {\rm chiral}} 
 + A_{n}^{[0]}\, ,
\equn
$$
where $A_{n}^{[0]}$ is the contribution from the complex scalar (or $\NeqZero$ matter multiplet) 
circulating in the loop. (Throughout we assume the use of a supersymmetry preserving
regulator~\cite{Siegel,StringBased,KST}.)

\vskip 15 pt 
\noindent
{\bf Recent Progress: }
Recently there has been a great deal of progress in calculating perturbative
amplitudes in compact forms:  this is key to our obtaining
gluino amplitudes.  Many of the techniques can be applied
directly to amplitudes involving particles other than gluons however
our philosophy will be to avoid such direct computations but rather to
exploit the gluonic amplitudes via the SWI. We have verified on
occasion our results using these methods - which we now review.

Progress in calculating amplitudes has been remarkable and varied.  At
tree level, inspired by the duality between topological string theory
and gauge theory~\cite{Witten:2003nn}, a reformulation of perturbation
theory in terms of MHV-vertices was proposed~\cite{CSW}. This promoted
the MHV amplitudes of eq.~(\ref{ParkeTaylor}) to the role of fundamental
building blocks in the perturbative expansion. By continuing legs
off-shell in a well specified manner these could be sewn together to
form other amplitudes.  This reformulation, although still lacking a
field theory proof, produces relatively compact expressions for tree
amplitudes. Although initially presented for purely gluonic amplitudes,
it has been successfully extended to other particle types
~\cite{CSW:matter,CSW:massive}.

In a different development,
a series of recursion relations for calculating tree
amplitudes have been postulated~\cite{Britto:2004ap}. These yield
compact expressions for gluonic tree amplitudes~\cite{Britto:2005dg}, the
six-point NMHV amplitudes involving fermions~\cite{Luo:2005rx}
and gravity amplitudes~\cite{Bedford:2005yy,Cachazo:2005ca,Bjerrum-Bohr:2005xx}.

Although impressive, progress in calculating tree amplitudes has
generally involved producing better forms for amplitudes which were
previously available (if only numerically). A much tougher but more
rewarding problem is to compute loop amplitudes for which much less is
known.  For one-loop amplitudes in gauge theory, full results for all
helicities and all particle types are only known for the
four-point~\cite{EllisSexton,KST} and five-point~\cite{FiveGluon,Bern:1994fz,Kunszt:1994tq,Glover:1996eh}
amplitudes.  Beyond five-point, the one-loop amplitudes are much
better understood within supersymmetric theories.

The MHV vertex approach has been
shown to extend to one-loop in ref.~\cite{Brandhuber:2004yw}, where the
one-loop $\NeqFour$ MHV amplitudes were computed and shown to be in complete
agreement with the results of ~\cite{BDDKa}, and in
refs~\cite{Quigley:2004pw,Bedford:2004py} where the $\NeqOne$ MHV
one-loop amplitudes were computed and shown to be in agreement with the results
of~\cite{BDDKb}.  Although the MHV vertex approach appears to work in
principle, the connection to the form of eqn.~(\ref{generalform2})
involves integration. Progress in evaluating the $c_i$ has been
more fruitful when employing methods which determine the coefficients using
algebraic equations. Techniques based on the structure of the
amplitude in twistor space can be used to give algebraic equations for
the box
coefficients~\cite{Cachazo:2004dr,BenaBernKosower,Britto:2004nj,Bidder:2004tx}
and techniques based on unitarity~\cite{BDDKa,BDDKb} can evaluate
the coefficients by evaluating the cuts of the amplitude.
For example, the box-coefficients must satisfy a coplanarity condition in twistor space,
$$
K_{ijkl} c^{NMHV }=
0\,,
\equn\label{EQcoplanar}
$$
where,
$$
K_{ijkl} =
\spa{i}.{j} \epsilon^{\dot a \dot b}
{ \partial \over \partial \tilde \lambda_k^{\dot a}  }
{ \partial \over \partial \tilde \lambda_l^{\dot b}  }
+{\rm perms}, \equn
$$
when the amplitude is expressed as a function of spinor variables 
$k_{a\dot a} =\lambda_a\tilde\lambda_a$. 
The coplanarity of the box-coefficients for the $\NeqFour$
amplitudes was shown in refs.~\cite{BDDK7,Britto:2004tx} and shown to extend 
to ${\cal N }< 4$ theories in 
in~\cite{BBDP,BBDP2005a}.  The box-coefficients we compute for amplitudes involving gluinos also satisfy~eq.(\ref{EQcoplanar}). 

These techniques have been very successful and  
results
include the recent computation of all $\NeqFour$ NMHV
one-loop amplitudes~\cite{BDDK7,BDKn} and 
various next-to-next-to-MHV (N$^2$MHV)
box coefficients~\cite{BrittoUnitarity}. 
An important
development, which enhances the power of the unitarity method, is
the observation by Britto, Cachazo and Feng~\cite{BrittoUnitarity} that
box integral coefficients can be obtained from generalised unitarity
cuts~\cite{Eden,GeneralizedCuts,BDDK7} by analytically 
continuing the massless corners of 
the quadruple cuts.  
The quadruple cuts give the 
box-coefficients as a product of four tree amplitudes,
$$
\eqalign{
c={ 1 \over 2 } \sum_{\cal S}
\biggl( \Atree(\ell_1,i_1,  & \ldots,i_2,\ell_2) \times
\Atree(\ell_2,i_3,\ldots,i_4,\ell_3)
\cr
& \hspace{2cm}\times \Atree(\ell_3,i_5,\ldots,i_6,\ell_4) \times
\Atree(\ell_4,i_7,\ldots,i_8,\ell_1) \biggr)\;.
\cr}
\equn\label{QuadCuts}
$$ 

\FIGURE[ht]{
\begin{picture}(160,100)(0,0)
\DashLine(50,73)(50,61){2}
\DashLine(50,39)(50,27){2}
\DashLine(27,50)(39,50){2}
\DashLine(61,50)(73,50){2}

\Line(30,30)(30,70)
\Line(70,30)(70,70)
\Line(30,30)(70,30)
\Line(70,70)(30,70)

\Line(30,70)(20,70)
\Line(30,70)(30,80)

\Line(70,30)(70,20)
\Line(70,30)(80,30)

\Line(70,70)(70,80)
\Line(70,70)(80,70)

\Line(30,20)(30,30)
\Line(20,30)(30,30)

\Text(30,10)[c]{$i_7$}
\Text(10,30)[c]{$i_8$}
\Text(25,25)[c]{$\bullet$}
\Text(70,10)[c]{$i_6$}
\Text(90,30)[c]{$i_5$}
\Text(70,90)[c]{$i_3$}
\Text(90,70)[c]{$i_4$}
\Text(30,90)[c]{$i_2$}
\Text(10,70)[c]{$i_1$}
\Text(75,25)[c]{$\bullet$}
\Text(75,75)[c]{$\bullet$}
\Text(25,75)[c]{$\bullet$}

\Text(20,52)[c]{$\ell_1$}
\Text(52,20)[c]{$\ell_4$}
\Text(80,52)[c]{$\ell_3$}
\Text(52,80)[c]{$\ell_2$}
\end{picture}
\label{QuadrupleCutFigure}
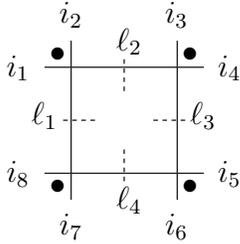
\caption{A quadruple cut of a $n$-point amplitude.  The dashed
lines represent the cuts.  The dots represent arbitrary numbers of
external line insertions.}
}
The sum is over all allowed intermediate configurations and
particle types~\cite{BrittoUnitarity} where the cut legs are frozen
in a specific manner. This formula could be used to compute the amplitudes
involving gluinos, however using the SWI produces
compact formulae in a straightforward manner. These formulae can be 
numerically compared
to the forms produced from~(\ref{QuadCuts}) as consistency checks.   

These techniques are also useful in calculating amplitudes in 
${\cal N} < 4$ theories \cite{Bidder:2004tx,BBDP, BBDP2005a,BrittoSQCD},  
although
these amplitudes are more complicated and contain
integral functions other than the box functions. 
Unfortunately, non-supersymmetric theories are not cut-constructible~\cite{BDDKb}, so the unitary
techniques are not immediately applicable, although progress is ongoing in 
this area~\cite{Bern:2005hs,Bern:2005ji}.

\section{Supersymmetric Ward Identities}   

Supersymmetric Ward Identities relate amplitudes with the
same helicity structure but with different external particles
types. The Ward identities can be obtained by acting with the
supersymmetry generator $Q$ on a string of operators, $z_i$, which has
vanishing vacuum expectation value.  
Typical choices are strings with
an odd number of fermionic operators.  
Since $Q$ annihilates the vacuum we obtain,
$$
0=  \biggl< \Bigl[ Q , \prod_{i} z_i \Bigr] \biggr>=
\sum_i \Bigl< z_1\cdots [Q,z_i]\cdots z_n \Bigr>\, . 
\equn
$$

\noindent
For $\NeqOne$ supersymmetry we can use the supersymmetry algebra, 
$$
\eqalign{
[Q(\eta) , g^{+} (p)] =- \Gamma^{+}(p,\eta) \gluinb^{+},\quad
[Q(\eta) , g^{-} (p)] =  \Gamma^{-}(p,\eta) \gluino^{-},
\cr
[Q(\eta) , \gluinb^{+} (p)] =- \Gamma^{-}(p,\eta) g^{+},\quad
[Q(\eta) , \gluino^{-} (p)] =  \Gamma^{+}(p,\eta) g^{-},
\cr}
\equn$$
where $g^{\pm}(p)$ is the operator creating a gluon of momentum $p$ and
the supersymmetry generator $Q(\eta)$  depends on a spinor parameter $\eta$. 
The $\Gamma^{\pm}$ are,
$$ 
\Gamma^{+}(p,\eta) \equiv \spb{\eta}.{p} \;\;, \;\;
\Gamma^{-}(p,\eta) \equiv \spa{p}.{\eta}. \equn
$$ 

Applying this to $\Ang'( g_1^-,g_2^- , \gluino_3^+,g_4^+,\ldots, g_n^+)$
(i.e. a string of glue creation operators with a single gluino creation operator) 
we obtain,
$$
\eqalign{
0=\spa1.{\eta} \Ang( \gluino_1^-,g_2^-,  \gluinb_3^+,g_4^+,\ldots, g_n^+)
+\spa2.{\eta} \Ang( g_1^-,\gluino_2^-, \gluinb_3^+,g_4^+,\ldots, g_n^+)
\cr
-\spa3.{\eta} \Ang( g_1^-,g_2^-,  g_3^+,g_4^+,\ldots, g_n^+)
\cr},
\equn$$ 
where we have used the fact that amplitudes with two fermions of the same helicity vanish. 
Choosing $\eta=1$, for example, gives,
$$
\Ang( g_1^-,\gluino_2^-, \gluinb_3^+,g_4^+,\ldots, g_n^+)=
{ \spa3.{1} \over \spa2.{1} } \Ang( g_1^-,g_2^-,  g_3^+,g_4^+,\ldots, g_n^+),
\equn$$
and we can thus obtain the 
MHV two-gluino amplitudes from the gluonic amplitude.

For NMHV amplitudes the SWI do not lead to such simple solutions:
applying the supersymmetry operator to $\Ang( g_1^-,g_2^-,g_3^- , \gluinb_4^+,g_5^+,\ldots, g_n^+)$
we obtain,
$$
\eqalign{
0=\spa1.{\eta} \Ang( \gluino_1^-,g_2^-,g_3^- , \gluinb_4^+,g_5^+,\ldots, g_n^+)
+\spa2.{\eta} \Ang( g_1^-,\gluino_2^-,g_3^- , \gluinb_4^+,g_5^+,\ldots, g_n^+)
\cr
+\spa3.{\eta} \Ang( g_1^-,g_2^-,\gluino_3^- , \gluinb_4^+,g_5^+,\ldots, g_n^+)
-\spa4.{\eta} \Ang( g_1^-,g_2^-,g_3^- , g_4^+,g_5^+,\ldots, g_n^+) \, . 
\cr}
\equn\label{SWInMHV}
$$ 
This system has rank 2, so it can only directly give two of the
amplitudes in terms of the other two. This relationship 
was used 
originally~\cite{Zoltan} to obtain the six point gluonic amplitude 
from the two-fermion amplitudes. By itself, this relationship 
does not allow us to
solve for the fermionic amplitudes unambiguously from the purely gluonic. 
However, when we
apply further constraints we will be able to obtain the fermionic amplitudes.

We can also consider 
$\NeqTwo$ Supersymmetric Ward Identities~\cite{SWI,Parke:1986jz}. 
Using supersymmetry generators 
$Q_a$, $a=1,2$, we have,
$$
\eqalign{
 [ Q_a (\eta) , g^{+} (p) ] &=- \Gamma^{+} (p,\eta) \gluinb_a^{+}, \quad
 [ Q_a (\eta) , g^{-} (p) ]  =  \Gamma^{-} (p,\eta) \gluino_a^{-},
\cr
[Q_a(\eta) , \gluinb_b^{+} (p)] &= -\Gamma^{-}(p,\eta) \delta_{ab}g^{+}- i \Gamma^{+}(p,\eta) \epsilon_{ab} \phi^{+},\cr
[Q_a(\eta) , \gluino_b^{-} (p)] &=  \Gamma^{+}(p,\eta) \delta_{ab}g^{-}+ i \Gamma^{-}(p,\eta) \epsilon_{ab} \phi^{-},
\cr
[ Q_a(\eta) ,\phi^{+} (p) ] &= - i \Gamma^{-} (p,\eta) \eps_{ab}\gluinb_b^{+}, \quad
[ Q_a(\eta) ,\phi^{-} (p) ]  = + i \Gamma^{+} (p,\eta) \eps_{ab}\gluino_b^{-}.
\cr}
\equn
$$
We will need to use these identities
to determine amplitudes involving scalars or two flavours of gluino.

\section{Tree Amplitudes}

In this section we demonstrate how to generate tree
amplitudes involving two gluinos from purely gluonic tree amplitudes and
then compare these to the known expressions obtained via recursion
relations~\cite{Luo:2005rx,BrittoSQCD} which themselves agree with the
Feynman diagram computations~\cite{Zoltan}.

\vskip 15 pt 
\noindent
{\bf Six-Point NMHV Tree Amplitudes} 

MHV amplitudes with two gluinos have been  discussed in section 2, in this section we 
consider NMHV amplitudes and compare our results with previous 
calculations~\cite{Zoltan,Luo:2005rx}.
For colour-ordered gluonic tree amplitudes
there are three independent NMHV helicity configurations.  When we consider
amplitudes with two fermions and four gluons there are considerably more
depending on the position of the two fermions.  This set can be 
reduced considerably using the $U(1)$ decoupling (or ``dual Ward'') identity~\cite{ManganoReview}.  
However, we will not explicitly use these identities since they do not extend 
to one-loop level, or more precisely, they have different implications.
We shall in this section restrict ourselves to adjoint fermions (gluinos).  

We first consider amplitudes derived from the gluonic amplitude,
$$
\eqalign{
\Ast( &g_1^-,g_2^-,g_3^- , g_4^+,g_5^+, g_6^+)
= 
\cr
&{i\BRi41{234}^3 \over t_{234} \spb2.3\spb3.4\spa5.6\spa6.1 \BRi25{234} }
+{i\BRi63{345}^3 \over t_{612} \spb6.1\spb1.2 \spa3.4\spa4.5 \BRi25{345}  } \, , 
\cr}
\equn\label{GluonTreeA}
$$
where, $\BRi{A}{B}{abc}\equiv
\la A^+\vert \slash\hskip -4pt k_a+\slash\hskip -4pt k_b+\slash\hskip -4pt k_c\vert B^+\ra
=\spb{A}.{a}\spa{a}.{B}+\spb{A}.{b}\spa{b}.{B}+\spb{A}.{c}\spa{c}.{B}$.
The amplitudes
involving two fermions which are
related to this purely gluonic amplitude can 
be obtained by conjugation, relabeling and {\it flipping} 
(i.e. using $A(1234556)=A(654321)$)  
from the following four,  
$$
\eqalign{
\Ast( \gluino_1^-,g_2^-,g_3^- , \gluinb_4^+,g_5^+, g_6^+),
\;\;\; & 
\Ast( g_1^-,\gluino_2^-,g_3^- , \gluinb_4^+,g_5^+,g_6^+),
\;\;\;
\cr 
\Ast(  g_1^-,g_2^-,\gluino_3^- , \gluinb_4^+,g_5^+, g_6^+),
\;\;\; & 
\Ast( g_1^-,\gluino_2^-,g_3^- , g_4^+,\gluinb_5^+,g_6^+)\, . 
\cr
}\equn
$$ 
The SWI~(\ref{SWInMHV}) relating the first three of
these amplitudes to the gluonic amplitude has rank 2 and hence, in
principle, is not sufficient to determine the fermionic amplitudes in
terms of the gluonic. 
However, when we utilise their inherent symmetries, we 
can unambiguously determine these fermionic amplitudes. 
The basic idea is to look for identities of the form,
$$
A\spa1.{\eta}
+B\spa2.{\eta}
+C\spa3.{\eta}
-D\spa4.{\eta}
=0 \, ,
\equn
$$
where the form of $D$ is motived by the terms in the numerator of
the compact expressions for the gluonic tree amplitudes. We shall search for
solutions where $A$, $B$ and $C$ are {\it polynomial} in the spinor invariants
$\spa{i}.{j}$ and $\spb{i}.{j}$, so that the gluino amplitudes are free 
from spurious singularities and poles. 

Equation~(\ref{GluonTreeA}) contains two terms which we examine individually.
Writing the second term as $\BRi63{612} X$ and focusing on the the $\BRi63{612}$ factor, the
Schouten identity yields,
$$
\eqalign{
\BRi63{612} \spa4.{\eta}
&=
-\BRi6{\eta}{612} \spa{3}.4 +
\BRi64{612} \spa3.{\eta}
\cr
&=
\BRi64{612} \spa3.{\eta}
-\spb6.1 \spa{3}.4\spa1.{\eta}
-\spb6.2 \spa{3}.4\spa2.{\eta}.
\cr}
\equn
$$
This implies that the following are solutions of the SWI,
$$
\eqalign{
\Ast( \gluino_1^-,g_2^-,g_3^- , \gluinb_4^+,g_5^+, g_6^+)
= &-\spb6.1 \spa{3}.4 X
\cr
\Ast( g_1^-,\gluino_2^-,g_3^- , \gluinb_4^+,g_5^+, g_6^+)
= &
-\spb6.2 \spa{3}.4 X
\cr
\Ast(  g_1^-,g_2^-,\gluino_3^- , \gluinb_4^+,g_5^+, g_6^+)
= &
\BRi64{612}  X
\cr
\Ast( g_1^-,g_2^-,g_3^- , g_4^+,g_5^+, g_6^+)
=  &
\BRi63{612}  X \, .
\cr}
\equn
$$
Similarly, writing the first term as $\BRi41{234}Y$ we find,
$$
\BRi41{234} \spa4.{\eta}=
\la 1 | K_{234} 4 | \eta \ra
=t_{234} \spa1.{\eta}
-\BRi21{234} \spa2.{\eta}
-\BRi31{234}\spa3.{\eta} \, , 
\equn
$$
which suggests a second solution to the SWI of the form,
$$
\eqalign{
\Ast( \gluino_1^-,g_2^-,g_3^- , \gluinb_4^+,g_5^+, g_6^+)
= & t_{234} Y
\cr
\Ast( g_1^-,\gluino_2^-,g_3^- , \gluinb_4^+,g_5^+, g_6^+)
= &
-\BRi21{234} Y
\cr
\Ast(  g_1^-,g_2^-,\gluino_3^- , \gluinb_4^+,g_5^+, g_6^+)
= &
-\BRi31{234} Y
\cr
\Ast( g_1^-,g_2^-,g_3^- , g_4^+,g_5^+, g_6^+)
=  &
\BRi41{234}  Y  \, . 
\cr}
\equn
$$
The two gluino tree amplitudes are thus,
$$
\eqalign{
\Ast(  g_1^-& ,g_2^-,\gluino_3^- , \gluinb_4^+,g_5^+, g_6^+)
= 
\cr
&
-{i\BRi41{234}^2\BRi31{234}  \over t_{234} \spb2.3\spb3.4\spa5.6\spa6.1 \BRi25{234} }
+{i\BRi63{612}^2 \BRi64{612}\over t_{612} \spb6.1\spb1.2 \spa3.4\spa4.5 \BRi25{612}  }
\cr
\Ast( g_1^-& ,\gluino_2^-,g_3^- , \gluinb_4^+,g_5^+,g_6^+)
=
\cr
& 
-{i\BRi41{234}^2 \BRi21{234} \over t_{234} \spb2.3\spb3.4\spa5.6\spa6.1 \BRi25{234} }
+{i\BRi63{612}^2 \spb2.6 \spa3.4
\over t_{612} \spb6.1\spb1.2 \spa3.4\spa4.5  \BRi25{612} }
\cr
\Ast( \gluino_1^- & ,g_2^-,g_3^- , \gluinb_4^+,g_5^+, g_6^+)
=
\cr
& 
 {i\BRi41{234}^2 t_{234} \over t_{234} \spb2.3\spb3.4\spa5.6\spa6.1 \BRi25{234} }
+{i\BRi63{612}^2 \spb1.6 \spa3.4 \over 
       t_{612} \spb6.1\spb1.2 \spa3.4\spa4.5 \BRi25{612} }  \, .
\cr}
\equn\label{EQFirstAmps}
$$
In principle there is some ambiguity in these solutions since the
coefficients of \break $\BRi63{612}$ and $\BRi41{234}$ are not unique:
$$
\eqalign{
\BRi63{612} &  X + \BRi41{234} Y \cr
& =\BRi63{612}\left( X+{Z \over \BRi63{612} } \right)
+\BRi41{234}\left(  Y-{Z \over\BRi41{234}}\right)
\cr}.
\equn
$$
However, by taking $X$ and $Y$ to be the values that appear in the gluon amplitudes
we do not introduce any of the unphysical singularities/poles that arise in the general
($Z\neq0$) case.
The remaining amplitude, $\Ast( g_1^-,\gluino_2^-,g_3^- , g_4^+,\gluinb_5^+,g_6^+)$, 
can be obtained from the SWI,
$$
\eqalign{
0=\spa1.{\eta} \Ast( \gluino_1^-,g_2^-,g_3^- , g_4^+,\gluinb_5^+,g_6^+)
+\spa2.{\eta} \Ast( g_1^-,\gluino_2^-,g_3^- , g_4^+,\gluinb_5^+, g_6^+)
\cr
+\spa3.{\eta} \Ast( g_1^-,g_2^-,\gluino_3^- , g_4^+,\gluinb_5^+, g_6^+)
-\spa5.{\eta} \Ast( g_1^-,g_2^-,g_3^- , g_4^+,g_5^+,g_6^+),
\cr}
\equn\label{SWInMHVb}
$$
which is obtained by acting with $Q$ on
$\Ast( g_1^-,g_2^-,g_3^- , g_4^+,\gluinb_5^+, g_6^+)$. 
Here we use the identities, 
$$
\eqalign{
\BRi63{612} \spa5.{\eta}
&=
\BRi65{612} \spa3.{\eta}
-\spb6.1 \spa{3}.5\spa1.{\eta}
-\spb6.2 \spa{3}.5\spa2.{\eta},
\cr
\BRi41{234} \spa5.{\eta}
&= \BRi45{234} \spa1.{\eta}
-\spb4.2  \spa1.5\spa2.{\eta}
-\spb4.3  \spa1.5\spa3.{\eta},
\cr}
\equn
$$
to obtain,
$$
\eqalign{
\Ast( g_1^-,\gluino_2^-,g_3^- , g_4^+,\gluinb_5^+, g_6^+)
& ={-i\BRi41{234}^2 \spb4.2 \spa1.5  \over t_{234} \spb2.3\spb3.4\spa5.6\spa6.1 \BRi25{234} }
\cr 
&
\hskip 2.0 truecm
 -{ i\BRi63{612}^2 \spb6.2\spa3.5  \over t_{612} \spb6.1\spb1.2 \spa3.4\spa4.5 \BRi25{612}  }.
\cr}
\equn
$$
This SWI also yields consistent but independent expressions for two of the amplitudes found previously.
For example,
$$
\eqalign{
\Ast( \gluino_1^-,g_2^-,g_3^- , g_4^+,\gluinb_5^+, g_6^+)
=&{i\BRi41{234}^2 \BRi45{234} \over t_{234} \spb2.3\spb3.4\spa5.6\spa6.1 \BRi25{234} }
\cr 
&
\hskip 2.0 truecm
-{ i\BRi63{612}^2 \spb6.1\spa3.5  \over t_{612} \spb6.1\spb1.2 \spa3.4\spa4.5 \BRi25{612}  }.
\cr}
\equn\label{EQSecondAmps}
$$
The expressions (\ref{EQFirstAmps}) and  (\ref{EQSecondAmps}) satisfy the consistency check,
$$
\Ast( \gluino_1^-,g_2^-,g_3^- , g_4^+,\gluinb_5^+, g_6^+)
=\biggl[
\Ast( g_1^-,\gluino_2^-,g_3^- ,\gluinb_4^+,g_5^+, g_6^+)
\biggr]\biggr|_{j\rightarrow j+3}.
\equn
$$
Thus we have a self-consistent set of six point, two gluino tree amplitudes for the helicity
configuration $(---+++)$.

Next we  consider the helicity configuration $(--+-++)$ and obtain 
two gluino amplitudes from the gluonic amplitude: 
$$
\eqalign{
&\Ast( g_1^-,g_2^-,g_3^+ , g_4^-,g_5^+,g_6^+)
=
  {
i\spa1.2^3 \spb5.6^3
\over t_{123} \spa2.3\spb4.5
\BRi41{123}\BRi63{123}}
\cr &
+
{i\BRi31{234}^4
\over
t_{234} \spb2.3\spb3.4\spa5.6\spa6.1
\BRi25{234}\BRi41{234}}
+
{i\BRi64{612}^4
\over
t_{345} \spb6.1\spb1.2\spa3.4\spa4.5
\BRi63{612}\BRi25{612}}.
\cr}
\equn
$$
Six amplitudes involving two gluinos are needed to generate all possibilities 
by relabeling, conjugation and flipping:
$$
\eqalign{
\Ast( \gluino_1^-,g_2^-,\gluinb_3^+ , g_4^-,g_5^+,g_6^+),
\;\;
\Ast( g_1^-,\gluino_2^-,\gluinb_3^+ , g_4^-,g_5^+,g_6^+),
\;\;
\Ast( g_1^-,g_2^-,\gluinb_3^+ , \gluino_4^-,g_5^+,g_6^+),
\cr
\Ast( \gluino_1^-,g_2^-,g_3^+ , g_4^-,\gluinb_5^+,g_6^+),
\;\;
\Ast( g_1^-,\gluino_2^-,g_3^+ , g_4^-,\gluinb_5^+,g_6^+),
\;\;
\Ast( \gluino_1^-,g_2^-,g_3^+ , g_4^-,g_5^+,\gluinb_6^+).
\cr}
\equn
$$
These are related to the gluonic amplitude 
via the three SWI,
$$
\eqalign{
0& =\spa1.{\eta} \Ast( \gluino_1^-,g_2^-,\gluinb_3^+ , g_4^-,g_5^+, g_6^+)
+\spa2.{\eta} \Ast( g_1^-,\gluino_2^-,\gluinb_3^+ , g_4^-, g_5^+,g_6^+)
\cr
 & +\spa4.{\eta} \Ast( g_1^-,g_2^-,\gluinb_3^+ , \gluino_4^-,g_5^+,g_6^+)
-\spa3.{\eta} \Ast( g_1^-,g_2^-,g_3^+ , g_4^-,g_5^+, g_6^+),
\cr
0& =\spa1.{\eta} \Ast( \gluino_1^-,g_2^-,g_3^+ , g_4^-,\gluinb_5^+, g_6^+)
+\spa2.{\eta} \Ast( g_1^-,\gluino_2^-,g_3^+ , g_4^-, \gluinb_5^+,g_6^+)
\cr
 & +\spa4.{\eta} \Ast( g_1^-,g_2^-,g_3^+ , \gluino_4^-,\gluinb_5^+,g_6^+)
-\spa5.{\eta} \Ast( g_1^-,g_2^-,g_3^+ , g_4^-,g_5^+, g_6^+),
\cr
0& =\spa1.{\eta} \Ast( \gluino_1^-,g_2^-,g_3^+ , g_4^-,g_5^+, \gluinb_6^+)
+\spa2.{\eta} \Ast( g_1^-,\gluino_2^-,g_3^+ , g_4^-, g_5^+,\gluinb_6^+)
\cr
 & +\spa4.{\eta} \Ast( g_1^-,g_2^-,g_3^+ , \gluino_4^-,g_5^+,\gluinb_6^+)
-\spa6.{\eta} \Ast( g_1^-,g_2^-,g_3^+ , g_4^-,g_5^+, g_6^+).
\cr}
\equn\label{cfbSWI}
$$ 
To solve the first of these, as before, we find two independent identities,
$$
\eqalign{
\BRi64{612} \spa{3}.{\eta}
&=
\BRi63{612}\spa{4}.{\eta}
-\spb1.6   \spa{3}.{4} \spa1.{\eta}
-\spb2.6 \spa{3}.{4} \spa2.{\eta},
\cr
\spa{3}.\eta \BRi31{234}
&=t_{234} \spa1.\eta
-\BRi21{234} \spa2.{\eta}
-\BRi41{234} \spa4.{\eta},
\cr}
\equn
$$
which give the following solutions to the SWI:
$$
\eqalign{
 \Ast( \gluino_1^-,g_2^-,\gluinb_3^+,g_4^-,g_5^+, g_6^+)
&=\spb6.1\spa3.4   X + t_{234} Y,
\cr
\Ast( g_1^-,\gluino_2^-,\gluinb_3^+,g_4^-,g_5^+, g_6^+)
&= \spb6.2\spa3.4  X -\BRi21{234} Y,
\cr
 \Ast( g_1^-,g_2^-,g_3^+ , g_4^-,g_5^+, g_6^+)
&= \BRi64{612}  X + \BRi31{234} Y,
\cr
\Ast( g_1^-,g_2^-,\gluinb_3^+,\gluino_4^-,g_5^+, g_6^+)
&=\BRi63{612}  X - \BRi41{234} Y.
\cr}
\equn
$$
We could  
rewrite the purely gluonic tree amplitude in the form \break $\BRi64{612}  X + \BRi31{234} Y $
by using the identity,
$$
{\spa1.2 \spb5.6 \over \BRi41{234} \BRi63{612} }
=
-
{\BRi31{234}
\over \BRi41{234} \BRi25{234} }
+
{\BRi64{612}
\over \BRi63{612} \BRi25{612} }.
\equn
$$
However, it is more convenient and in line with our philosophy of not
generating extra poles
to use the Schouten identity to produce,
$$
\spa3.{\eta}\spa1.2  \spb5.6= \spa1.{\eta}\spa3.2  \spb5.6+
\spa2.{\eta}\spa1.3  \spb5.6.
\equn
$$
Whether we rearrange to use two identities or use three, we obtain the 
same solutions, 
$$
\eqalign{
&\Ast( \gluino_1^-, g_2^-,\gluinb_3^+,g_4^-,g_5^+, g_6^+)
= {-
i\spa1.2^2  \spa2.3\spb5.6^3 
\over t_{123} \spa2.3\spb4.5 \BRi41{123} \BRi63{123}
}
\cr
&+
{i\BRi31{234}^3 t_{234}
\over
t_{234} \spb2.3\spb3.4\spa5.6\spa6.1
\BRi25{234}\BRi41{234}}
+
{ i\BRi64{612}^3 \spb6.1\spa3.4
\over
t_{345} \spb6.1\spb1.2\spa3.4\spa4.5
\BRi63{612}\BRi25{612}}
\cr
&\Ast( g_1^-,\gluino_2^-,\gluinb_3^+,g_4^-,g_5^+,g_6^+)
=
 {
i\spa1.2^2 \spa1.3 \spb5.6^3
\over t_{123} \spa2.3\spb4.5 \BRi41{123} \BRi63{123} }
\cr
&+
{ -i\BRi31{234}^3 \BRi21{234}
\over
t_{234} \spb2.3\spb3.4\spa5.6\spa6.1
\BRi25{234}\BRi41{234}}
+
{ i\BRi64{612}^3 \spb6.2\spa3.4
\over
t_{345} \spb6.1\spb1.2\spa3.4\spa4.5
\BRi63{612}\BRi25{612}}
\cr
&\Ast( g_1^-,g_2^-,\gluinb_3^+,\gluino_4^-,g_5^+, g_6^+)
=
\cr
&
{ -i \BRi31{234}^3  \BRi41{234}
\over
t_{234} \spb2.3\spb3.4\spa5.6\spa6.1
\BRi25{234}\BRi41{234}}
+
{ i\BRi64{612}^3 \BRi63{612}
\over
t_{345} \spb6.1\spb1.2\spa3.4\spa4.5
\BRi63{612}\BRi25{612} }
\cr}
\equn
$$
The remaining two amplitudes can be obtained similarly.  

For the final gluonic configuration,
$$
\eqalign{
&\Ast( g_1^-,g_2^+,g_3^-,g_4^+,g_5^-,g_6^+)
={i \BRi25{123}^4
\over
t_{123} \spb1.2\spb2.3\spa4.5\spa5.6
\BRi14{123}\BRi36{123}}
\cr &
+
{ i\BRi63{234}^4
\over
t_{234} \spa2.3\spa3.4\spb5.6\spb6.1
\BRi52{234}\BRi14{234}}
+
{ i\BRi41{345}^4
\over
t_{345} \spa6.1\spa1.2\spb3.4\spb4.5
\BRi36{345}\BRi52{345}},
\cr}
\equn
$$
there are two independent amplitudes involving two gluinos,
$$
\Ast( \gluino_1^-,\gluinb_2^+,g_3^-,g_4^+,g_5^-,g_6^+)\;\;\;,
\Ast( g_1^-,\gluinb_2^+,g_3^-,g_4^+,\gluino_5^-,g_6^+),
\equn
$$ 
which we can obtain from the SWI, 
$$
\eqalign{
0=\spa1.{\eta} \Ast( \gluino_1^-,\gluinb_2^+,g_3^- , g_4^+,g_5^-, g_6^+)
+\spa3.{\eta} \Ast( g_1^-,\gluinb_2^+,\gluino_3^- , g_4^+,g_5^-, g_6^+)
\cr 
-\spa2.{\eta} \Ast( g_1^-,g_2^+,g_3^- , g_4^+,g_5^-, g_6^+)
+\spa5.{\eta} \Ast( g_1^-,\gluinb_2^+,g_3^- , g_4^+,\gluino_5^-, g_6^+).
\cr}
\equn\label{cfcSWI}
$$ 
We  solve this using  the identities,
$$
\eqalign{
\BRi25{123} \spa2.{\eta}  &= t_{123}  \spa5.{\eta}
-\BRi15{123} \spa1.{\eta}
-\BRi35{123}\spa3.{\eta}
\cr
\BRi63{234} \spa2.{\eta} &= \BRi62{234} \spa3.{\eta}
+\spa2.3 \spb5.6 \spa5.{\eta}
-\spa2.3 \spb6.1 \spa1.{\eta}
\cr
\BRi41{345}  \spa2.{\eta}  &= \BRi42{345} \spa1.{\eta}
+\spb3.4 \spa1.2 \spa3.{\eta}
-\spb4.5 \spa1.2 \spa5.{\eta},
\cr}
\equn
$$
giving the tree amplitudes, 
$$
\eqalign{
&\Ast( \gluino_1^-,\gluinb_2^+,g_3^-,g_4^+,g_5^-,g_6^+)
={ -i\BRi25{123}^3 \BRi15{123}
\over
t_{123} \spb1.2\spb2.3\spa4.5\spa5.6
\BRi14{123}\BRi36{123} }
+
\cr&
{ -i\BRi63{234}^3\spa2.3 \spb6.1
\over
t_{234} \spa2.3\spa3.4\spb5.6\spb6.1
\BRi52{234}\BRi14{234}}
+
{ i\BRi41{345}^3 \BRi42{345} 
\over
t_{345} \spa6.1\spa1.2\spb3.4\spb4.5
\BRi36{345}\BRi52{345}}
\cr
&\Ast( g_1^-,\gluinb_2^+,\gluino_3^-,g_4^+,g_5^-,g_6^+)
={ -i \BRi25{123}^3  \BRi35{123} 
\over
t_{123} \spb1.2\spb2.3\spa4.5\spa5.6
\BRi14{123}\BRi36{123}}
+
\cr&
{ i\BRi63{234}^3\BRi62{234}
\over
t_{234} \spa2.3\spa3.4\spb5.6\spb6.1
\BRi52{234}\BRi14{234}}
+
{ i\BRi41{345}^3\spb3.4 \spa1.2
\over
t_{345} \spa6.1\spa1.2\spb3.4\spb4.5
\BRi36{345}\BRi52{345}}
\cr 
&\Ast(  g_1^-,\gluinb_2^+,g_3^-,g_4^+,\gluino_5^-,g_6^+)
= {i \BRi25{123}^3 t_{123} 
\over
t_{123} \spb1.2\spb2.3\spa4.5\spa5.6
\BRi14{123}\BRi36{123}}
+
\cr&
{ i\BRi63{234}^3\spa2.3 \spb5.6
\over
t_{234} \spa2.3\spa3.4\spb5.6\spb6.1
\BRi52{234}\BRi14{234}}
+
{ -i\BRi41{345}^3\spb4.5 \spa1.2
\over
t_{345} \spa6.1\spa1.2\spb3.4\spb4.5
\BRi36{345}\BRi52{345}}
\cr}
\equn
$$

The six-point two-quark amplitudes have been computed
previously~\cite{Zoltan} and can be obtained in compact expressions
using recursion relations~\cite{Luo:2005rx}.  Our results for adjacent
gluinos match these exactly - demonstrating that by respecting the
symmetries and factorisation structures of the amplitudes one can use
the SWI to generate the correct results.

\section{Six Point One Loop NMHV Amplitudes with two Gluinos}
 \label{TwoGluinoSection}                                                                               
The SWI apply to all orders in perturbation theory, so we
can apply our technique  to one-loop amplitudes. Furthermore,
 $\NeqFour$ one-loop amplitudes can be expressed as sums of box integrals with
rational coefficients~\cite{BDDKa}.  Since the box integrals are an
independent set of functions the SWI for these amplitudes will apply
box by box.

For the six-point, one loop, NMHV amplitudes the only 
types of box contributing are the 
``two-mass-hard'' and one-mass boxes. 
These appear in certain very specific combinations:
$$
\eqalign{
  \Wsix{i}\ &\equiv\ \Fone{i} + \Fone{i+3}                 
+ \Fhard{2;i+1} + \Fhard{2;i+4} 
\cr
    \ &=\ -{1\over2\e^2} \sum_{j=1}^6
         \left( { \mu^2 \over -s_{j,j+1} } \right)^\e
 \ -\ \ln\left({-t_{i,i+1,i+2} \over -s_{i,i+1}}\right)
      \ln\left({-t_{i,i+1,i+2} \over -s_{i+1,i+2}}\right)\cr
 &\quad
  \ -\ \ln\left({-t_{i,i+1,i+2} \over -s_{i+3,i+4}}\right)
      \ln\left({-t_{i,i+1,i+2} \over -s_{i+4,i+5}}\right)
 \ +\ \ln\left({-t_{i,i+1,i+2} \over -s_{i+2,i+3}}\right)
      \ln\left({-t_{i,i+1,i+2} \over -s_{i+5,i}}\right)\cr
 &\quad
 \ +\ {1\over 2}\ln\left({-s_{i,i+1} \over -s_{i+3,i+4}}\right)
         \ln\left({-s_{i+1,i+2} \over -s_{i+4,i+5}}\right)
 \ +\ {1\over 2}\ln\left({-s_{i-1,i} \over -s_{i,i+1}}\right)
         \ln\left({-s_{i+1,i+2} \over -s_{i+2,i+3}}\right)\cr
  &\quad
 \ +\ {1\over 2}\ln\left({-s_{i+2,i+3} \over -s_{i+3,i+4}}\right)
         \ln\left({-s_{i+4,i+5} \over -s_{i+5,i}}\right)
 \ +\ {\pi^2\over3}\ . \cr}
\equn\label{EQWsix}
$$
There are only three independent $\Wsix{i}$ since $\Wsix{i+3}=\Wsix{i}$. 
The  $\Wsix{i}$ have several rather special features which will extend to
amplitudes involving fermions. Firstly, even though the integral functions
individually contain dilogarithms,  these drop out of the $\Wsix{i}$.
Secondly, the IR singularities take the rather simple form,
$$
 \Wsix{i}=
-{3\over\e^2} + \half \sum_{j=1}^6  
{ \ln(   -s_{j,j+1})  \over \eps} 
+O (\eps^0), 
\equn
$$
which leads to the sum of the coefficients of the $\Wsix{i}$ 
being proportional to  the tree amplitude.

The first set of amplitudes we shall consider are based on the
gluonic amplitude,
$$
\Asf(1^-,2^-,3^-,4^+,5^+,6^+)\ =\
  \rg \LB B_1\,\Wsix1+B_2\,\Wsix2+B_3\,\Wsix3\RB,
\equn
$$
where,
$$
\eqalign{
  B_1\ &=\ B_0 \equiv i { (t_{123})^3
  \over \spb1.2\spb2.3\spa4.5\spa5.6\
\BRi14{123} \BRi36{123} } , \cr
  B_2\ &=\ \left({ \BRi41{234} 
         \over t_{234} } \right)^4  \ B_+
       + \left({ \spa2.3\spb5.6 \over t_{234} } \right)^4
            \ B_+^\cc , \cr
  B_3\ &=\ \left({ \BRi63{345} 
         \over t_{345} } \right)^4  \ B_-
       + \left({ \spa1.2\spb4.5 \over t_{345} } \right)^4
            \ B_-^\cc  , \cr}
\equn
$$
and,
$$
B_+ = B_0 |_{j \rightarrow j+1} 
\;\; , \;\;\;
B_- = B_0 |_{j \rightarrow j-1} \ , 
\equn
$$
where the operation ${}^\dagger$ implies $\spb{i}.{j} \leftrightarrow \spa{j}.{i}$. 
This amplitude has two symmetries,
$$
\eqalign{
S_1\;\; : \Asf(1^-,2^-,3^-,4^+,5^+,6^+)\ & = 
\left[ \Asf(1^-,2^-,3^-,4^+,5^+,6^+) \right]^\dagger_{j\rightarrow j+3},
\cr
S_2\;\; : \Asf(1^-,2^-,3^-,4^+,5^+,6^+)\ & = 
\left[ \Asf(1^-,2^-,3^-,4^+,5^+,6^+) \right]^\dagger_{j \rightarrow 6-j},
\cr}
\equn
$$
which impose constraints on the coefficients.  Under $S_1$, 
$W_i\rightarrow W_i$ so we have,
$$
S_1: B_i \longrightarrow B_i
\equn
$$
whereas under $S_2$, $W_1 \rightarrow W_1$ and $W_2 \leftrightarrow W_3$
so that
$$
S_2: 
B_1 \longrightarrow B_1,
 \;\;\;  
B_2 \leftrightarrow B_3.
\equn
$$ 
The coefficients clearly satisfy these conditions 
when we note that $B_0$ itself satisfies,
$$
S_1 : B_0 \longrightarrow B_0,
\;\;\;
S_2 : B_0 \longrightarrow B_0.
\equn
$$
Applying $S_i$ to the gluino amplitudes provides a set of consistency conditions
that enable us to resolve the ambiguities that arise in solving the SWI.

As for the tree amplitudes, we can generate all 
the possible two-gluino amplitudes from a minimal set of 
four by conjugation, relabeling and flipping.
These gluino amplitudes have a subset of the invariances of the gluonic amplitudes. Specifically,  
$A( g_1^-,\gluino_2^-,g_3^- ,g_4^+,\gluinb_5^+,g_6^+)$ is invariant under $S_1$ and $S_2$,
while $A(\gluino_1^-,g_2^-,g_3^- , \gluinb_4^+,g_5^+, g_6^+)$ 
is only invariant under $S_1$, 
$A( g_1^-,g_2^-,\gluino_3^- , \gluinb_4^+,g_5^+, g_6^+)$
is only  invariant under $S_2$ 
and
$A(g_1^-,\gluino_2^-,g_3^- , \gluinb_4^+,g_5^+,g_6^+)$ 
is invariant under neither. 

For this helicity configuration the SWI are (\ref{SWInMHV}) and (\ref{SWInMHVb}).
To solve for $B_1$ we need identities involving $\spa4.\eta$ and $\spa5.\eta$. These are,
$$
\eqalign{\;\;\;t_{123}\spa4.\eta
=&
\BRi14{123} \spa1.\eta
+\BRi24{123} \spa2.\eta
+\BRi34{123} \spa3.\eta,
\cr
t_{123}\spa5.\eta
=&\BRi15{123} \spa1.\eta
+\BRi25{123} \spa2.\eta
+\BRi35{123} \spa3.\eta.
\cr}
\equn\label{B1ids}
$$
We can check that these equations are consistent with the symmetries $S_i$: 
if we have solutions,
$$
\eqalign{
A \spa4.\eta& = B  \spa1.\eta+C \spa2.\eta +D\spa3.\eta,
\cr
A' \spa5.\eta& = B'  \spa1.\eta+C' \spa2.\eta +D'\spa3.\eta,\cr}
\equn
$$
then we must have,
$$
\eqalign{
S_1 : (B/A) \rightarrow (B/A) , 
\;\;\;
S_2 : (D/A)  \rightarrow( D/A) ,
\;\;\;
S_i : (C'/A) \rightarrow (C'/A). 
\;\;\;
\cr}
\equn
$$
The coefficients in (\ref{B1ids}) clearly satisfy these constraints.
Thus we have solutions, 
$$ 
\eqalign{ 
B_1({\gluino_1^-,g_2^-,g_3^- , \gluinb_4^+,g_5^+,g_6^+}) 
& =
{i(t_{123})^{2}\BRi14{123}  \over \spb1.2\spb2.3
\spa4.5\spa5.6\BRi14{123}\BRi36{123}}, 
\cr 
B_1({g_1^-,\gluino_2^-,g_3^-,\gluinb_4^+,g_5^+,g_6^+})
& = 
{i(t_{123})^{2}\BRi24{123}  \over
\spb1.2\spb2.3 \spa4.5\spa5.6\BRi14{123}\BRi36{123}} ,
\cr 
B_1({g_1^-,g_2^-,\gluino_3^- ,\gluinb_4^+,g_5^+,g_6^+}) 
& = 
{i(t_{123})^2\BRi34{123}  \over
\spb1.2\spb2.3 \spa4.5\spa5.6\BRi14{123}\BRi36{123}}, 
\cr 
B_1({g_1^-,\gluino_2^-,g_3^- ,g_4^+,\gluinb_5^+,g_6^+}) 
& = 
{i(t_{123})^2  \BRi25{123} \over
\spb1.2\spb2.3 \spa4.5\spa5.6\BRi14{123}\BRi36{123}} .
\cr} \equn
$$

To solve for the first three $B_2$'s we use the identities,
$$
\eqalign{
\BRi41{234}\spa4.\eta   =& 
t_{234} \spa1.{\eta}
-\BRi21{234} \spa2.{\eta}
-\BRi31{234}\spa3.{\eta},
\cr
\spa2.3 \spa4.\eta   =& \spa4.3 \spa2.\eta+\spa2.4 \spa3.\eta.
\cr}
\equn
$$
Which give solutions,
$$ 
\eqalign{ 
B_2({\gluino_1^-,g_2^-,g_3^- , \gluinb_4^+,g_5^+,g_6^+}) 
& =
 \left({ \BRi41{234}^3 
         \over t_{234}^3 } \right)  \ B_+,
\cr 
B_2({g_1^-,\gluino_2^-,g_3^-,\gluinb_4^+,g_5^+,g_6^+})
& = 
 \left({ - \BRi41{234}^3 \BRi21{234}
         \over t_{234}^4 } \right)  \ B_+
       + \left({ \spa2.3^3\spa4.3\spb5.6^4 \over t_{234}^4 } \right)
            \ B_+^\cc ,
\cr 
B_2({g_1^-,g_2^-,\gluino_3^- ,\gluinb_4^+,g_5^+,g_6^+}) 
& = 
 \left({ -\BRi41{234}^3 \BRi31{234}
         \over t_{234}^4 } \right)  \ B_+
       + \left({ \spa2.3^3\spa2.4\spb5.6^4 \over t_{234}^4 } \right)
            \ B_+^\cc .
\cr}
\equn
$$ 
The absence of a second term from the first coefficient is
consistent with the observation that this box-coefficient does not
have a singlet term when we consider two-particle cuts in the
$t_{234}$ channel.  (This observation would naturally lead us to an identity 
that does not  involve $\spa1.\eta $)

For the final $B_2$ box coefficient there are three identities we might 
use:
$$
\eqalign{
\BRi41{234} \spa5.{\eta}
=& \BRi45{234} \spa1.{\eta}
-\spb4.2  \spa1.5\spa2.{\eta}
-\spb4.3  \spa1.5\spa3.{\eta},
\cr
\spa2.3\spb5.6 \spa5.\eta
=&-\spa2.3\spb1.6\spa1.\eta +\BRi63{234}\spa2.\eta- \BRi62{234}\spa3.\eta,
\cr
\spa2.3\spa5.\eta
=&\spa5.3 \spa2.\eta+\spa2.5 \spa3.\eta.
\cr}
\equn
$$
Of these, only the first two have the correct behavior under $S_i$.
Using these identities we find,
$$
\eqalign{ 
B_2({g_1^-,\gluino_2^-,g_3^- ,g_4^+,\gluinb_5^+,g_6^+}) 
& = 
{ -\BRi41{234}^3 \spb4.2\spa1.5
         \over t_{234}^4 }   B_+
       + { \spa2.3^3\spb5.6^3\BRi63{234} \over t_{234}^4 } 
             B_+^\cc ,
\cr} \equn
$$ 
which has the appropriate symmetries. This pair of identities also lead to the same forms for 
the other $B_2$ coefficients obtained previously.

For the $B_3$ coefficients,  the identities,
$$
\eqalign{
\BRi63{345} \spa4.{\eta}
=&
\spb6.1 \spa{3}.4\spa1.{\eta}
+\spb6.2 \spa{3}.4\spa2.{\eta}
+\BRi64{345} \spa3.{\eta},
\cr
\spa1.2 \spb4.5 \spa4.\eta   =&
-\BRi52{345} \spa1.\eta +\BRi51{345} \spa2.\eta
 -\spa1.2 \spb3.5 \spa3.\eta,  
\cr
\BRi63{345} \spa5.{\eta}
=&
+\spb6.1 \spa{3}.5\spa1.{\eta}
+\spb6.2 \spa{3}.5\spa2.{\eta}
+\BRi65{345} \spa3.{\eta},
\cr
\spa1.2\spb4.5 \spa5.\eta
=& -\spa1.2\spb4.3 \spa3.\eta  
+\BRi42{345} \spa1.\eta  
-\BRi41{345} \spa2.\eta, 
\cr}
\equn
$$
give the following solutions with the correct symmetries under $S_1$,
$$ 
\eqalign{ 
B_3({\gluino_1^-,g_2^-,g_3^- , \gluinb_4^+,g_5^+,g_6^+}) 
& =
 \left({ \BRi63{345}^3\spa3.4\spb6.1  
         \over t_{345}^4  } \right)   B_-
       + \left({ -\spa1.2^3\spb4.5^3 \BRi52{345} \over t_{345}^4 } \right)
             B_-^\cc ,
\cr 
B_3({g_1^-,\gluino_2^-,g_3^-,\gluinb_4^+,g_5^+,g_6^+})
& = 
 \left({ \BRi63{345}^3 \spa3.4\spb6.2   
         \over t_{345}^4  } \right)   B_-
       + \left({ \spa1.2^3\spb4.5^3 \BRi51{345} \over t_{345}^4  } \right)
             B_-^\cc ,
\cr 
B_3({g_1^-,g_2^-,\gluino_3^- ,\gluinb_4^+,g_5^+,g_6^+}) 
& = 
 \left({ \BRi63{345}^3\BRi64{345}
         \over t_{345}^4 } \right)  B_- 
       + \left({ -\spa1.2^4\spb4.5^3\spb3.5  \over t_{345}^4  } \right)
             B_-^\cc,
\cr 
B_3({g_1^-,\gluino_2^-,g_3^- ,g_4^+,\gluinb_5^+,g_6^+}) 
& = 
 \left({ \BRi63{345}^3\spb6.2\spa3.5
         \over t_{345}^4 } \right)   B_- 
       + \left({ -\spa1.2^3\spb4.5^3 \BRi41{345} \over t_{345}^4  } \right)
             B_-^\cc .
\cr} \equn
$$ 
Comparing these with the $B_2$ coefficients we see that  the $S_2$ symmetry is also satisfied.

We can obtain the gluino amplitudes with helicity configurations
$(--+-++)$ and $(-+-+-+)$ in a similar manner, i.e. by  finding polynomial solutions
to the SWI based on the gluonic amplitudes that respect the
symmetries of the amplitudes.
We have verified numerically that these 
expressions agree with those obtained using quadruple cuts~(\ref{QuadCuts}).  
These coefficients are collected in the appendix and 
are available in Mathematica
format to download.

There are straightforward relationships between the 
box coefficients and tree amplitudes.  
By necessity the IR divergences must be of the form,
$$
\sim  \Atree \times \sum_i { \ln (s_{ii+1} ) \over \eps } \ .
\equn
$$
The box-coefficients must then satisfy, 
$$
B_1+B_2+B_3  = 2\Atree \ .
\equn\label{EQIR1}
$$ 
It can be checked numerically that this is true by comparison with
the tree amplitudes of section~4.  The expressions for the tree amplitudes actually
correspond to a subset of the terms comprising $B_i$.  There are five
terms in each $B_1+B_2+B_3$ expression.  Two of these correspond
exactly to the tree amplitudes of section~4 whereas the other three
give an alternate, not trivially related, expression for the tree
amplitude.  For example taking the amplitude
$\Ast(  g_1^-,g_2^-,\gluino_3^- , \gluino_4^+,g_5^+, g_6^+)$
we find,
$$
\eqalign{
\Ast =& {(t_{123})^3\BRi34{123}  \over
\spb1.2\spb2.3 \spa4.5\spa5.6\BR14\BR36} 
       + \left({ \spa2.3^3\spa2.4\spb5.6^4 \over t_{234}^4 } \right)
            \ B_+^\cc ,
 \cr
&  \hskip 3.0 truecm      
+ \left({ -\spa1.2^4\spb4.5^3\spb3.5  \over t_{345}^4  } \right)
            \ B_-^\cc
\cr
=&  \left({ -\BR41^3 \BR31
         \over t_{234}^4 } \right)  \ B_+
+ \left({ \BR63^3\BR64
         \over t_{345}^4 } \right)  \ B_-   \ , 
\cr}
\equn
$$ 
the second expression being that for the tree amplitude of
section~4 and reference~\cite{Luo:2005rx}.
These relationships mirror very closely the behaviour of the gluon scattering amplitudes:
it was  the observation that the box coefficients reproduced the tree amplitudes in 
simple compact forms~\cite{BDKn,Roiban:2004ix} that led to the recursion 
relationships for tree amplitudes~\cite{Britto:2004ap}.

The twistor structure of the box coefficients is also rather simple: 
all the box-coefficients satisfy coplanarity constraints,
$$
K_{abcd}  B_i  =0 \ . 
\equn
$$ 
In fact this is satisfied by each of the terms within $B_i$
individually.

\section{Amplitudes With More Than Two Fermions} 

We can use the SWI to obtain amplitudes involving four or more gluinos of the same flavour from
those involving two gluinos. In the six-point case the tree amplitudes involving four and six 
fermions have been computed directly~\cite{Gunion:1986zh,Xu:1986xb} and also using 
recursion relations~\cite{Luo:2005B}. 

If we consider 
$n$-point NMHV amplitudes with negative helicities on legs $m_i$,
applying the $\NeqOne$ supersymmetry operator to, 
$$
\Ang( g_1^+,\ldots , g_{m_1}^-, \ldots   g_{m_2}^-, \ldots \gluino_{m_3}^-, 
\ldots \gluinb_r^+ \ldots \gluinb_s^+\ldots , g_n^+),
\equn
$$
gives the SWI,
$$
\eqalign{
0= \spa{m_1}.\eta 
A_{n}^{m_1,m_3;r,s}
+
\spa{m_2}.\eta
A_{n}^{m_2,m_3;r,s}
-\spa{r}.\eta
A_{n}^{m_3;s}
-\spa{s}.\eta
A_{n}^{m_3;r}
\cr},
\equn
$$
where we define, 
$$
\eqalign{
A_{n}^{m_2,m_3;r,s}
\equiv \Ang( g_1^+,\ldots , g_{m_1}^-, \ldots   \gluino_{m_2}^-, \ldots ,\gluino_{m_3}^-, 
\ldots \gluinb_r^+ \ldots \gluinb_s^+\ldots , g_n^+),
\cr
A_{n}^{m_3;r}
\equiv \Ang( g_1^+,\ldots , g_{m_1}^-, \ldots   g_{m_2}^-, \ldots ,\gluino_{m_3}^-, 
\ldots \gluinb_r^+ \ldots g_s^+\ldots , g_n^+).\cr}
\equn
$$
This rank two system can be used to solve for the four fermion amplitudes in terms of the amplitudes with two fermions. 
For example choosing $\eta=m_1$ gives,
$$
\eqalign{
A_{n}^{m_2,m_3;r,s}
= 
{ \spa{r}.{m_1} \over \spa{m_2}.{m_1} }
A_{n}^{m_3;r}
+{ \spa{s}.{m_1}  \over \spa{m_2}.{m_1}  }
A_{n}^{m_3;s}. 
\cr}
\equn
$$ 
Since we have used the $\NeqOne$ SWI, all of the fermions in this amplitude
have the same flavour.

To obtain amplitudes with six gluinos we apply the supersymmetry
operator to,
$$
\Ang( g_1^+,\ldots , g_{m_1}^-, \ldots   \gluino_{m_2}^-,\gluino_{m_3}^-, \ldots \gluinb_r^+ \ldots \gluinb_s^+\ldots , 
 \gluinb_t^+\ldots g_n^+),
\equn
$$
giving the SWI, 
$$
\eqalign{
0=& 
\spa{m_1}.\eta 
A_{n}^{m_1,m_2,m_3;r,s,t}
-\spa{r}.\eta
A_{n}^{m_2,m_3;s,t}
-\spa{s}.\eta
A_{n}^{m_2,m_3;r,t}
-\spa{t}.\eta
A_{n}^{m_2,m_3;r,s},
\cr}
\equn
$$ 
which allows us to express the six fermion amplitude in terms of
four fermion amplitudes. For example, choosing $\eta=r$,
$$
A_{n}^{m_1,m_2,m_3;r,s,t}
= { \spa{s}.r \over \spa{m_1}.r }
A_{n}^{m_2,m_3;r,t}
+{\spa{t}.r \over \spa{m_1}.r}
A_{n}^{m_2,m_3;r,s}.
\equn
$$
Again the fermions are all of the same
flavour.  These relations are exact to all orders in perturbation
theory in any supersymmetric theory.  

For amplitudes involving two fermion flavours we must be precise about
which theory we are describing and in particular whether our theory
contains scalars. Supersymmetric amplitudes with two flavours of
fermions must include at least one scalar.  For ${\cal N} \geq 2$ (and
indeed for $\NeqOne$ with adjoint matter) the fermions have  Yukawa couplings
to the scalars which simultaneously change both the flavour and
the helicity of the fermions. Such Yukawa couplings do not contribute to tree amplitudes 
with two gluinos, but they can contribute to amplitudes with four
gluinos of two different flavours.

In $\NeqTwo$ we can generate a SWI by 
applying $Q_2$ to,
$$
\Anz( g_1^+,\ldots , g_{m_1}^-, \ldots   g_{m_2}^-,\ldots,
\gluino_{m_3}^{1-}, \ldots \gluinb_{r}^{1+} \ldots \gluinb_{s}^{2+}\ldots , g_n^+).
\equn
$$
We obtain,
$$
\eqalign{
0= & 
\spa{m_1}.\eta
\Anz( g_1^+,\ldots ,\gluino_{m_1}^{2-}, \ldots   g_{m_2}^-,\ldots, 
\gluino_{m_3}^{1-}, 
\ldots \gluinb_{r}^{1+} \ldots \gluinb_{s}^{2+}\ldots , g_n^+)
\cr
+& 
\spa{m_2}.\eta
\Anz( g_1^+,\ldots ,g_{m_1}^-, \ldots   \gluino_{m_2}^{2-},\ldots,
\gluino_{m_3}^{1-}, \ldots \gluinb_{r}^{1+} \ldots \gluinb_{s}^{2+}\ldots , g_n^+)
\cr
-i& 
\spa{m_3}.\eta
\Anz( g_1^+,\ldots , g_{m_1}^-, \ldots   g_{m_2}^-,\ldots,
\phi_{m_3}^-, \ldots \gluinb_{r}^{1+} \ldots \gluinb_{s}^{2+}\ldots , g_n^+)
\cr
-& 
\spa{s}.\eta
\; \Anz( g_1^+,\ldots , g_{m_1}^-, \ldots   g_{m_2}^-,\ldots,
\gluino_{m_3}^{1-}, \ldots \gluinb_{r}^{1+} \ldots g_{s}^+\ldots , g_n^+),
\cr}
\equn\label{FourFermSWI}
$$ 
which can be used to determine the two flavour, four fermion amplitude 
in terms of a two fermion amplitude we have already calculated and 
a scalar-fermion-fermion amplitude which we discuss in the next section.

\section{Amplitudes Involving Scalars}

As noted above, for ${\cal N} \geq 2$ the fermions have Yukawa couplings to the
scalars which simultaneously change both the flavour
and the helicity of the fermion. 
At tree level, this vertex implies that amplitudes of the form,
$$
\Ant( \phi^-, \gluino^{1+} , \gluino^{2+} , g^{\pm}, \ldots, g^{\pm} ),
\equn
$$ 
need not vanish.  In fact, there are non-vanishing MHV tree amplitudes
of this form as may be seen in the expression of ~\cite{Nair}. 
These amplitudes will
appear in the SWI and must not be discarded.

In an $\NeqTwo$ theory there are two flavours of 
gluino, $\gluino^i$.  Acting with $Q_2$ on, 
$$
\Anz(  \gluino_1^{1-},g_2^-,g_3^- ,\phi_4^+,g_5^+,\ldots, g_n^+),
\equn
$$
gives,
$$
\eqalign{0=&
-i \spa1.{\eta} 
\Anz( \phi_1^-,g_2^-,  g_3^- ,\phi_4^+,g_5^+,\ldots, g_n^+)
\cr
+&
\spa2.{\eta} \Anz( \gluino_1^{1-}, \gluino_2^{2-},  g_3^- , \phi_4^+,g_5^+,\ldots, g_n^+)
\cr
+&\spa3.{\eta}\Anz( \gluino_1^{1-},g_2^-,  \gluino_3^{2-} , \phi_4^+,g_5^+,\ldots, g_n^+)
\cr
+&i
\spa4.{\eta} \Anz( \gluino_1^{1-},g_2^-,  g_3^- , \gluinb_4^{1+},g_5^+,\ldots, g_n^+).
\cr}
\equn\label{ScalarSWI}
$$   
To solve this we need to find polynomial expressions of the form,
$$
0=  iA \spa1.{\eta} 
+B\spa2.{\eta} 
+C\spa3.{\eta} 
-iD\spa4.{\eta}. 
\equn\label{SWIidentityScalar}
$$
Given such solutions, there will be relationships between the {\it individual terms}
of the two gluino and two scalar amplitudes of the form,
$$
A_n^{term}( \phi_1^-,g_2^-,  g_3^- ,\phi_4^+,g_5^+,\ldots, g_n^+)
=
\left( {  A \over D} \right)  
A_n^{term}( \gluino_1^{1-},g_2^-,  g_3^- , \gluinb_4^{1+},g_5^+,\ldots, g_n).
\equn
$$
If the appropriate solutions to~(\ref{SWIidentityScalar}) are the same as those used
to obtain the two-gluino amplitudes, then the scalar terms will
be of the form,
$$
A_n^{term}( \phi_1^-,g_2^-,  g_3^- ,\phi_4^+,g_5^+,\ldots, g_n^+)
=
\left( {  A \over D} \right)^2  
A_n^{term}( g_1^-,g_2^-,  g_3^- , g_4^+,g_5^+,\ldots, g_n).
\equn
$$
For gluonic amplitudes of the form,
$$
A^{gluon}_n = \sum_i X_i,
\equn
$$
we might expect amplitudes containing a pair of particles of spin $h$ 
to have the form,
$$
A_n^{h-{\rm pair}} = \sum_i  (a_i)^{2-2h} X_i,
\equn
$$
where $h=1$ for gluons, $h=1/2$ for fermions and $h=0$ for scalars. 
Such structures are apparent in tree amplitudes as can be seen in the 
results of~\cite{Luo:2005rx,BrittoSQCD}.  
For example, we can generalise our two gluino tree amplitude for the helicity
configuration\break
 $(---+++)$ to give, 
$$
\eqalign{
\Asz (H_1^-,g_2^-,  g_3^- ,H_4^+,g_5^+,g_6^+)
=& 
\left( { t_{234} \over\BRi41{234} } \right)^{2-2h} 
{i\BRi41{234}^3 \over t_{234} \spb2.3\spb3.4\spa5.6\spa6.1 \BRi25{234} }
\cr&
+\left( { \spb1.6 \spa3.4\over \BRi63{612} } \right)^{2-2h}  
{ i\BRi63{612}^3  \over t_{612} \spb6.1\spb1.2 \spa3.4\spa4.5 \BRi25{612} },  
\cr}
\equn
$$ 
where $H$ represent a gluon for $h=1$, a gluino for $h=1/2$, a scalar for $h=0$
and an anti-gluino for $h=-1/2$. 
Such formulae are extremely useful when computing
one-loop amplitudes using cuts (see for example~\cite{BBDP2005a,BrittoSQCD}).  

This behaviour extends to the coefficients of the one-loop box
functions and we give expressions for the box-functions for two
scalars in the appendix.  We have checked numerically for a
representative sample that the box-coefficients thus
obtained match those obtained via quadruple cuts.

Once we have the two gluino and two scalar amplitudes, the SWI~(\ref{ScalarSWI}) gives
amplitudes such as,
$$
A_n( \gluino_1^{1-}, \gluino_2^{2-},  g_3^- , \phi_4^+,g_5^+,g_6^+),
\equn
$$
directly.  Given these amplitudes, the two flavour, four gluino amplitudes 
can be obtained directly from~(\ref{FourFermSWI}).

\section{Conclusions}

Recently there has been much progress in computing one-loop amplitudes
with external gluons in compact forms, where the factorisation
structure is simple and manifest.  In this paper we have shown how to
use Supersymmetric Ward Identities, together with the inherent
symmetries of the amplitude, to generate one-loop amplitudes where the
external particles are gluinos or adjoint scalars from these compact
gluonic expressions.  In particular, we have calculated all the
six-point $\NeqFour$ NMHV amplitudes involving two gluinos or scalars.
The amplitudes with four or six gluinos (of a single flavour) have
been given as linear combinations of the two gluino amplitudes.
Although these results are specific to supersymmetric theories and
adjoint fermions, they do reduce the amount of computation required to
obtain results in non-supersymmetric theories with fundamental quarks.
We also expect SWI to facilitate the calculation of loop scattering amplitudes 
in supergravity theories (see appendix B).

Organising Yang-Mills amplitudes in terms of helicity structure,
particle type, colour and supersymmetry has helped enormously in
understanding the structure of interactions in Yang-Mills theory
and is the framework within which progress has occurred.  However,
phenomenologically, most of the quantum numbers which organise the
amplitude are not observable in collider experiments.  Consequently,
the list of simple amplitudes required to compute an experimental
quantity is rather long. The use of symmetries, such as SWI, to generate amplitudes 
without explicit computation is very helpful in this context.

\begin{acknowledgments}
We thank Emil Bjerrum-Bohr and  Harald Ita for useful
discussions.  This work was supported by a PPARC rolling grant.
SB would like to thank PPARC for a research studentship.

\end{acknowledgments}

\begin{appendix}

\section{Summary of One-Loop Two Gluino and Two Scalar Six-Point Amplitudes
in $\NeqFour$ SYM}
\label{appGluino}

The amplitudes for the $\NeqFour$ theory are all of the form,
$$
\Asf(1,2,3,4,5,6)\ =\
 \rg  \LB c_1\,\Wsix1+c_2\,\Wsix2+c_3\,\Wsix3\RB,
\equn
$$
with the coefficients $c_i$ 
depending on the helicity and type of the six particles.  
This combination of box functions is given explicitly in eq.~(\ref{EQWsix}).  
The amplitudes will have one particle denoted $H$ and a second denoted by $\bar H$. 
Again, $H$ will denote either a scalar or $\gluino^{\pm}$.  
The amplitudes are obtained
using the specific values of $h$ 
as defined in table~1.   
\begin{table}[h]
\hbox{
\def\tend{\cr \noalign{ \hrule}}
\vbox{\offinterlineskip
{
\hrule
\halign{
       &  \vrule#
\strut        &\strut\hfil #\hfil\vrule
 \strut        &\strut\hfil #\hfil\vrule
 \strut         &\strut\hfil #\hfil\vrule
       \cr
height15pt  &$\;\; H\;\; $  & $\;\; \bar H\;\; $     &
$\;\; h$   & \tend
height20pt & $g^- $  & $g^+$     &
1   & \tend
height20pt & $\gluino^- $  & $\gluinb^+$     &
1/2   & \tend
height20pt & $\phi^-$  & $\phi^+$     &
0   & \tend
height20pt & $\gluino^+ $  & $\gluinb^-$     &
-1/2   & \tend
}
}
}
}
\nobreak
\caption[]{The values of $h$  for the choices of external particle $H$.
\label{AmplitudesD8}
\smallskip}
\end{table}

We express the box coefficients in terms of $B_0$ and $B_{\pm}$
and their conjugates where,
$$
B_0=i { (t_{123})^3
  \over \spb1.2\spb2.3\spa4.5\spa5.6\
\BR14 \BR36 },
$$
and 
$$
B_+ = B_0 |_{j \rightarrow j+1} 
\;\; , \;\;\;
B_- = B_0 |_{j \rightarrow j-1}. 
\equn
$$

For amplitudes with helicity configuration $(---+++)$ we denote the
$c_i$ in the purely gluonic case by $B_i$,
$$
A_{6}^{N=4}(1^-,2^-,3^-,4^+,5^+,6^+)\ =\
\rg   \LB B_1\,\Wsix1+B_2\,\Wsix2+B_3\,\Wsix3\RB,
\equn
$$
where,
$$
\eqalign{
  B_1\ &=\ B_0  , \cr
  B_2\ &\equiv B_2^A+B_2^B =\ \left({ \BRi41{234} 
         \over t_{234} } \right)^4  \ B_+
       + \left({ \spa2.3\spb5.6 \over t_{234} } \right)^4
            \ B_+^\cc , \cr
  B_3\ &\equiv B_3^A+B_3^B =\ \left({ \BRi63{345} 
         \over t_{345} } \right)^4  \ B_-
       + \left({ \spa1.2\spb4.5 \over t_{345} } \right)^4
            \ B_-^\cc  . \cr}
\equn
$$

For ease of presentation we shall denote the
box-coefficients with fermions/scalars as $B_i^{ab}$ when legs $a$ and $b$ are the $H$ and
$\bar H$ particles.  The solutions for the $B_i^{ab}$ for gluinos were derived in
section~\ref{TwoGluinoSection} and we present them here again in a
form that also gives the two scalar amplitudes. For the four independent configurations,
$(ab)=(14),$ $(24),$ $(34)$ or $(25),$ 
we find, 
$$ 
\eqalign{ 
B_1^{14}
& =
\left( { \BRi14{123}\over t_{123}} \right)^{2-2h}  B_0,
\cr 
B_1^{24}
& = 
\left( { \BRi24{123}\over t_{123}} \right)^{2-2h}  B_0,
\cr 
B_1^{34}
& = 
\left( { \BRi34{123}\over t_{123}} \right)^{2-2h}  B_0,
\cr 
B_1^{25}
& = 
\left( { \BRi25{123}\over t_{123}} \right)^{2-2h}  B_0,
\cr} 
\equn
$$
$$ 
\eqalign{ 
B_2^{14}
 =&
 \left({ t_{234} \over \BRi41{234}   } \right)^{2-2h} 
B_2^A,
\cr 
B_2^{24}
 =& 
 \left({ -\BRi21{234}
         \over  \BRi41{234} } \right)^{2-2h}  
B_2^A
+ \left({\spa4.3 \over   \spa2.3} \right)^{2-2h}  
B_2^B,
\cr 
B_2^{34}
 =& 
 \left({  -\BRi31{234}
         \over \BRi41{234} } \right)^{2-2h}  
B_2^A
       + 
\left({ \spa2.4 \over \spa2.3 } \right)^{2-2h}  
B_2^B,
\cr
B_2^{25} 
= &
\left( {  -\spb4.2\spa1.5
         \over \BRi41{234} }  \right)^{2-2h} 
B_2^A
       + 
\left( {  \BRi63{234} \over \spa2.3\spb5.6  } \right)^{2-2h} 
B_2^B,
\cr}
\equn
$$
$$ 
\eqalign{ 
B_3^{14}
& =
 \left({ \spa3.4\spb6.1  
         \over \BRi63{345}  } \right)^{2-2h}   
B_3^A
       + 
\left({  -\BRi52{345} \over \spa1.2\spb4.5 } \right)^{2-2h}
B_3^B,
\cr 
B_3^{24}
& = 
 \left({ \spa3.4\spb6.2   
         \over  \BRi63{345}  } \right)^{2-2h}  
B_3^A 
       + 
\left({  \BRi51{345} \over  \spa1.2\spb4.5} \right)^{2-2h}
B_3^B,
\cr 
B_3^{34} 
& = 
 \left({ \BRi64{345}
         \over \BRi63{345} } \right)^{2-2h} 
 B_3^A 
       + 
\left(- {\spb3.5  \over   \spb4.5 } \right)^{2-2h}
B_3^B,
\cr 
B_3^{25}
& = 
 \left({\spb6.2\spa3.5
         \over  \BRi63{345} } \right)^{2-2h}  
B_3^A  
       + 
\left({ -\BRi41{345} \over   \spa1.2\spb4.5 } \right)^{2-2h}
B_3^B.
\cr} \equn
$$ 

Next we have amplitudes with helicity structure $(--+-++)$.  
In the purely gluonic case, the amplitude is symmetric under, 
$$
S_1 : \Asf(1,2,3,4,5,6) \longrightarrow 
[ \Asf(6,5,4,3,2,1) ]^\dagger.
\equn
$$
In this case we denote the coefficients of the $W_6^{(i)}$ by $D_i$. These are given by,
$$
\eqalign{
D_1\equiv D_1^A+D_1^B=\ 
  \left({ \BRi34{123}
         \over t_{123} } \right)^4   
B_0
       + 
\left({ \spa1.2\spb5.6 \over t_{123} } \right)^4
             B_0^\cc  ,
\cr
D_2 \equiv D_2^A+D_2^B =
\left({ \BRi31{234} \over t_{234} } \right)^4      
\ B_{+} 
+ 
\left({ \spa2.4\spb5.6 \over t_{234} } \right)^4
\ B_{+}^{\cc},
\cr
D_3\equiv D_3^A+D_3^B =\ 
\left( { \BRi64{345}         \over t_{345} } \right)^4 
\ B_{-}       
+ 
\left({ \spa1.2\spb3.5 \over t_{345} } \right)^4             
\ B_{-}^\cc .
\cr}
\equn
$$

As above,
we denote the coefficients of amplitudes with particle $a$ of type
$H$ and particle $b$ of type $\bar H$ by $D_i^{ab}$.  For this helicity
configuration there are six independent possibilities:
$$
(ab)= (13),(23),(43),(15),(25),(16).
\equn
$$
These six box-coefficients are constrained by the system of three SWI~(\ref{cfbSWI}). 
In solving these we must find solutions which satisfy,
$$
\eqalign{
S_1: D_1^{ab} \longrightarrow D_1^{ab} \;\;\; (ab)=(34),(25),(16),
\cr
S_1: D_2^{ab} \leftrightarrow D_3^{ab} \;\;\; (ab)=(34),(25),(16).
\cr}
\equn
$$ 

The identities that give amplitudes with the appropriate symmetries are:
$$
\eqalign{
\BRi34{123}\spa{3}.{\eta} 
=&
t_{123}\spa{4}.{\eta} - \BRi14{123}\spa{1}.{\eta} - \BRi24{123}\spa{2}.{\eta}
,\cr
\spa1.2\spb5.6\spa{3}.{\eta} 
=&
\spa1.3\spb5.6\spa{2}.{\eta} + \spa3.2\spb5.6\spa{1}.{\eta}
,\cr
\BRi31{234}\spa{3}.{\eta} 
=&
t_{234}\spa{1}.{\eta} - \BRi21{234}\spa{2}.{\eta} - \BRi41{234}\spa{4}.{\eta}
,\cr
\spa2.4\spb5.6\spa{3}.{\eta}
=&
\spa3.4\spb5.6\spa{2}.{\eta} + \spa2.3\spb5.6\spa{4}.{\eta}
,\cr
\BRi64{345}\spa{3}.{\eta}
=&
\spa4.3\spb6.1\spa{1}.{\eta} + \spa4.3\spb6.2\spa{2}.{\eta} + \BRi63{345}\spa{4}.{\eta}
,\cr
\spa1.2\spb3.5\spa{3}.{\eta}
=&
-\BRi52{345}\spa{1}.{\eta} + \BRi51{345}\spa{2}.{\eta} - \spa1.2\spb4.5\spa{4}.{\eta}
,\cr}
\equn
$$
$$
\eqalign{
\BRi34{123}\spa{5}.{\eta}
=&
\BRi35{123}\spa{4}.{\eta} - \spb3.1\spa4.5\spa{1}.{\eta} - \spb3.2\spa4.5\spa{2}.{\eta}
,\cr
\spa1.2\spb5.6\spa{5}.{\eta} 
=&
-\spa1.2\spb4.6\spa{4}.{\eta} - \BRi61{123}\spa{2}.{\eta} + \BRi62{123}\spa{1}.{\eta}
,\cr
\BRi31{234}\spa{5}.{\eta}
=&
\BRi35{234}\spa{1}.{\eta} - \spa1.5\spb3.2\spa{2}.{\eta} - \spa1.5\spb3.4\spa{4}.{\eta}
,\cr
\spa2.4\spb5.6\spa{5}.{\eta} 
=&
-\spa2.4\spb1.6\spa{1}.{\eta} - \BRi62{234}\spa{4}.{\eta} + \BRi64{234}\spa{2}.{\eta}
,\cr
\BRi64{345}\spa{5}.{\eta} 
=& \BRi65{345}\spa{4}.{\eta} + \spa4.5\spb6.1\spa{1}.{\eta} + \spa4.5\spb6.2\spa{2}.{\eta}
,\cr
\spa1.2\spb3.5\spa{5}.{\eta} 
=& - \spa1.2\spb3.4\spa{4}.{\eta} - \BRi31{345}\spa{2}.{\eta}+\BRi32{345}\spa{1}.{\eta}
,\cr}
\equn
$$

$$
\eqalign{
\BRi34{123}\spa{6}.{\eta}
=&
\BRi36{123}\spa{4}.{\eta} - \spb3.1\spa4.6\spa{1}.{\eta} - \spb3.2\spa4.6\spa{2}.{\eta}
,\cr
\spa1.2\spb5.6\spa{6}.{\eta} 
=&
\spa1.2\spb4.5\spa{4}.{\eta} + \BRi51{123}\spa{2}.{\eta} - \BRi52{123}\spa{1}.{\eta}
,\cr
\BRi31{234}\spa{6}.{\eta}
=&
\BRi36{234}\spa{1}.{\eta} - \spa1.6\spb3.2\spa{2}.{\eta} - \spa1.6\spb3.4\spa{4}.{\eta}
,\cr
\spa2.4\spb5.6\spa{6}.{\eta} 
=&
\spa2.4\spb1.5\spa{1}.{\eta} +\BRi52{234}\spa{4}.{\eta} - \BRi54{234}\spa{2}.{\eta}
,\cr
\BRi64{345}\spa{6}.{\eta} 
=& -t_{345}\spa{4}.{\eta} - \BRi14{345}\spa{1}.{\eta} - \BRi24{345}\spa{2}.{\eta}
,\cr
\spa1.2\spb3.5\spa{6}.{\eta} 
=& - \spb3.5\spa6.1\spa{2}.{\eta}-\spb3.5\spa2.6\spa{1}.{\eta}
.\cr}
\equn
$$
The box coefficients are then given by,
$$
\eqalign{
  D_1^{13}
= & 
\ 
\left(-{ \BRi14{123}
         \over \BRi34{123} } \right)^{2-2h}   
D_1^A
       + 
\left({ \spa3.2\over \spa1.2  } \right)^{2-2h}   
D_1^B
,\cr
D_1^{23} 
= &
 \left(-{ \BRi24{123}
         \over \BRi34{123} } \right)^{2-2h}    
D_1^A       + 
\left({ \spa1.3\over \spa1.2 } \right)^{2-2h}
D_1^B
,\cr
D_1^{43} 
=  &
 \ 
\left( {t_{123}\over \BRi34{123} } \right)^{2-2h}   
D_1^A
,\cr
D_1^{15} 
= &
 \ 
\left( {\spa4.5 \spb1.3 \over \BRi34{123} } \right)^{2-2h}   
D_1^A   
    + 
\left({ \BRi62{123}\over \spa1.2\spb5.6 } \right)^{2-2h}
D_1^B
,\cr
D_1^{25} 
=  &
\left( { \spa4.5 \spb2 .3\over \BRi34{123} } \right)^{2-2h}   
D_1^A+
\left(-{ \BRi61{123}\over \spa1.2\spb5.6 } \right)^{2-2h}
D_1^B
,\cr
D_1^{16} 
=  & \ 
\left( -{\spb3.1\spa4.6 \over \BRi34{123} } \right)^{2-2h}   
D_1^A
       + 
\left({ - \BRi52{123}\over \spa1.2\spb5.6 } \right)^{2-2h}
D_1^B
,\cr}
\equn
$$
$$
\eqalign{
  D_2^{13} 
&=
 \ 
\left({ t_{234}\over  \BRi31{234} } \right)^{2-2h}  
 D_2^A 
,\cr
D_2^{23} 
&=
\ 
\left(-{ \BRi21{234} \over  \BRi31{234}} \right)^{2-2h}      
 D_2^A 
+ 
\left({ \spa3.4 \over \spa2.4 } \right)^{2-2h}            
 D_2^B
,\cr
D_2^{43}
&=
\ 
\left(-{\BRi41{234} 
         \over\BRi31{234}  } \right)^{2-2h}        
 D_2^A       
+ \left({ \spa2.3 \over \spa2.4 } \right)^{2-2h}            
 D_2^B  
,\cr
D_2^{15}
&=
\ 
\left({\BRi35{234} 
         \over\BRi31{234}  } \right)^{2-2h}        
D_2^A
+ \left(-{ \spb1.6 \over \spb5.6 } \right)^{2-2h}            
D_2^B
,\cr
D_2^{25}
&=
\ 
\left({\spb2.3\spa1.5 
         \over\BRi31{234}  } \right)^{2-2h}        
D_2^A       
+ \left({ \BRi64{234} \over \spa2.4\spb5.6 } \right)^{2-2h}            
D_2^B 
,\cr
D_2^{16}
&=
\ 
\left({\BRi36{234} 
         \over\BRi31{234}  } \right)^{2-2h}        
D_2^A       
+ \left(-{\spb5.1 \over \spb5.6 } \right)^{2-2h}            
D_2^B 
,\cr}
\equn
$$

$$
\eqalign{
  D_3^{13} 
= & 
\ 
\left( -{  \spb6.1\spa3.4  \over \BRi64{345} } \right)^{2-2h}       
D_3^A       
+ 
\left(-{ \BRi52{345} \over \spa1.2\spb3.5  } \right)^{2-2h}
D_3^B
,\cr
 D_3^{23}
= & 
\ \left(-{  \spb6.2\spa3.4  \over \BRi64{345}  } \right)^{2-2h}    
D_3^A       
+ 
\left({ \BRi51{345} \over \spa1.2\spb3.5} \right)^{2-2h}   
D_3^B
,\cr
D_3^{43} 
= &
\ 
\left({  \BRi63{345}  \over \BRi64{345}  } \right)^{2-2h} 
D_3^A 
+
\left(-{ \spb4.5 \over \spb3.5 } \right)^{2-2h} 
D_3^B
, \cr
 D_3^{15}
= & 
\ \left({  \spb6.1\spa4.5  \over \BRi64{345}  } \right)^{2-2h}    
D_3^A       
+ 
\left({ \BRi32{345} \over \spa1.2\spb3.5} \right)^{2-2h}   
D_3^B 
,\cr
 D_3^{25}
= & 
\ \left({  \spb6.2\spa4.5  \over \BRi64{345}  } \right)^{2-2h}    
D_3^A       
+ 
\left(-{ \BRi31{345}\over \spa1.2\spb3.5} \right)^{2-2h}   
D_3^B 
,\cr
 D_3^{16}
= & 
\ \left(-{  \BRi14{345}  \over \BRi64{345}  } \right)^{2-2h}    
D_3^A       
+ 
\left({ \spa6.2 \over \spa1.2} \right)^{2-2h}   
D_3^B 
.\cr}
\equn
$$

Next we have amplitudes with helicity structure $(-+-+-+)$.  
The purely gluonic amplitude is symmetric under, 
$$
\eqalign{
S_1 &: \Asf(1,2,3,4,5,6) \longrightarrow 
 \Asf(1,2,3,4,5,6)|_{j \rightarrow j+2} 
,\cr
S_2 &: \Asf(1,2,3,4,5,6) \longrightarrow 
 [ \Asf(1,2,3,4,5,6)|_{j \rightarrow j+1}]^\dagger 
.\cr}
\equn
$$  
In this case we denote the coefficients of $W_6^{(i)}$ in the purely gluonic
case by $G_i$. These are given by,
$$
\eqalign{
G_1\equiv G_1^A+G_1^B=\ 
\left({ \BRi25{123}
         \over t_{123} } \right)^4
\ B_0
       + 
\left({ \spa1.3\spb4.6 \over t_{123} } \right)^4
\ B_0^\cc \ ,
\cr
G_2 \equiv G_2^A+G_2^B =
\left({ \BRi63{234}
 \over t_{234} } \right)^4   
\ B_{+}^\cc 
+
\left({ \spa5.1 \spb2.4 \over t_{234} } \right)^4
            \ B_{+} ,
\cr
G_3\equiv G_3^A+G_3^B =\ 
 \left({ \BRi41{345}
         \over t_{345} } \right)^4      
\ B_{-}^\cc 
+
\left({\spa3.5\spb6.2 \over t_{345} } \right)^4 B_{-}.
\cr}
\equn
$$
Although there are only two independent configurations with two gluinos in this case,
we present results for all the two gluino amplitudes appearing in 
the SWI~(\ref{cfcSWI}).  Amplitudes with the correct symmetries are produced by applying
the following identities to the SWI:
$$
\eqalign{
\BRi25{123}\spa{2}.{\eta} 
&= 
t_{123}\spa{5}.{\eta} - \BRi15{123}\spa{1}.{\eta} - \BRi35{123}\spa{3}.{\eta}
,\cr
\spa1.3\spb4.6\spa{2}.{\eta} 
&=
\spa2.3\spb4.6\spa{1}.{\eta} + \spa1.2\spb4.6\spa{3}.{\eta}
,\cr 
\BRi63{234}\spa{2}.{\eta}
&=
-\spb6.1\spa2.3\spa{1}.{\eta} 
+ \BRi62{234}\spa{3}.{\eta} + \spb5.6\spa2.3\spa{5}.{\eta}
,\cr
\spa5.1\spb2.4\spa{2}.{\eta} 
&=
-\spa5.1\spb3.4\spa{3}.{\eta} + \BRi45{234}\spa{1}.{\eta} - \BRi41{234}\spa{5}.{\eta}
,\cr
\BRi41{345}\spa{2}.{\eta}
&=
\BRi42{345}\spa{1}.{\eta} 
+ \spb3.4\spa1.2\spa{3}.{\eta} -\spb4.5\spa1.2\spa{5}.{\eta}
,\cr
\spa3.5\spb6.2\spa{2}.{\eta} 
&=
-\spa3.5\spb6.1\spa{1}.{\eta} - \BRi65{345}\spa{3}.{\eta} + \BRi63{345}\spa{5}.{\eta}
.\cr}
\equn
$$

The amplitudes are:
$$
\eqalign{
 G_1^{12} 
=&
\left(-{ \BRi15{123} 
         \over \BRi25{123}  } \right)^{2-2h}    
G_1^A
+ 
\left({   \spa2.3 \over \spa1.3  } \right)^{2-2h} 
G_1^B ,
\cr
G_1^{32}
=&  
\left(-{  \BRi35{123}
         \over \BRi25{123} } \right)^{2-2h}        
G_1^A
+ 
\left({\spa1.2   \over  \spa1.3 } \right)^{2-2h} 
G_1^B,
\cr
G_1^{52} 
=&  \left({ t_{123}  \over\BRi25{123}  } \right)^{2-2h} 
G_1^A
,\cr}
\equn
$$

$$
\eqalign{
 G^{12}_2
=&  
\left(-{ \spa2.3 \spb6.1
 \over  \BRi63{234}} \right)^{2-2h}     
G_2^A 
+ \left({ \BRi45{234}\over \spa5.1\spb2.4 } \right)^{2-2h}
G_2^B
,\cr
G^{32}_2
=&  
\left({ \BRi62{234}  \over \BRi63{234}   } \right)^{2-2h} 
G_2^A
+ \left(-{ \spb3.4 \over \spb2.4 } \right)^{2-2h}
G_2^B
,\cr
G^{52}_2
=&  
\left({ \spa2.3 \spb5.6  \over \BRi63{234} } \right)^{2-2h}  
G_2^A
+\left(-{\BRi41{234}\over \spa5.1\spb2.4 } \right)^{2-2h}  
G_2^B
,\cr}
\equn
$$
$$
\eqalign{
 G^{12}_3
=&
\left({ \BRi42{345}
         \over \BRi41{345} } \right)^{2-2h}    
G_3^A 
+ \left(-{\spb6.1 \over \spb6.2 } \right)^{2-2h} G_3^B
,\cr
G^{32}_3
=&  
\left({ \spb3.4 \spa1.2 \over \BRi41{345} } \right)^{2-2h}  
 G_3^A
+ \left(-{ \BRi65{345} \over \spa3.5\spb6.2 } \right)^{2-2h}  
G_3^B
,\cr
G^{52}_3
=&  
\left(-{ \spb4.5 \spa1.2   \over \BRi41{345} } \right)^{2-2h}      
G_3^A
+ \left({ \BRi63{345}  \over  \spa3.5\spb6.2 } \right)^{2-2h}  
G_3^B
.\cr}
\equn
$$

The six-point box-coefficients have been explicitly checked by 
numerically evaluating the
quadruple cuts.
The functional forms of the 
six-point coefficients $B_i^{ab},D_i^{ab}$ and $G_i^{ab}$ are available 
in Mathematica format at: 

http://pyweb.swan.ac.uk/$\ \tilde{}$ dunbar/SWAT430.html.

\section{Graviton Scattering Amplitudes}

Some of the earliest applications of SWI were to 
graviton scattering amplitudes~\cite{SWI}.  
The MHV amplitudes involving different particle 
types obey relationships very similar to the 
Yang-Mills case,
$$
M(g^-_1, H^-_2, H^+_3 , g_4, \cdots , g_n^+ )
= \left( 
{ \spa1.3 \over \spa1.2 }\right)^{4-2h}
M(g^-_1, g^-_2, g^+_3 , g_4, \cdots , g_n^+ ),
\equn
$$
where $h$ now runs over the helicities of the $\NeqEight$ supergravity multiplet,
i.e.  from $h=-2$ to $h=+2$.  

If we solve the SWI for NMHV amplitudes, 
we might again expect to find amplitudes of the form,
$$
\sum_i   (X_i)^{4-2h}  \times C_i.
\equn
$$
Examination of the six-point NMHV tree amplitude~\cite{Cachazo:2005ca,Bjerrum-Bohr:2005xx} reveals precisely this structure
and it can also be found in the coefficients of loop amplitudes~\cite{BeBbDu,Bjerrum-Bohr:2005xx,BDDPR}.

\end{appendix}

\end{document}